\documentclass{iopart}
\usepackage{iopams,bm,cite,color,graphicx,subfigure,braket,mathptmx,bm,color,url}

\eqnobysec

\usepackage{graphicx,array,enumitem,epstopdf,boxedminipage,bm,color}

\usepackage[colorlinks=true,linkcolor=blue,citecolor=blue,urlcolor =blue]{hyperref}
\usepackage{mleftright}

\newcommand{\nn}{\nonumber\\}

\newcommand{\be}{\begin{equation}}
\newcommand{\ee}{\end{equation}}
\newcommand{\bea}{\begin{eqnarray}}
\newcommand{\eea}{\end{eqnarray}}

\newcommand{\abssq}[1]{\left| #1 \right|^2}

\newcommand{\eqref}[1]{\mbox{Eq.~(\ref{#1})}}

\begin{document}

\title{Catch and release of propagating bosonic field with non-Markovian giant atom}

\author{Luting Xu$^{1}$ and Lingzhen Guo$^{1,2}$}
\address{$^1$Center for Joint Quantum Studies and Department of Physics, School of Science, Tianjin University, Tianjin 300072, China }
\address{$^2$Max Planck Institute for the Science of Light (MPL), Staudtstrasse 2, 91058 Erlangen, Germany}
\ead{lingzhen\_guo@tju.edu.cn}
\vspace{10pt}
\begin{indented}
\item[]\today
\end{indented}

\begin{abstract}
The non-Markovianity of physical systems is considered to be a valuable resource that has potential applications to quantum information processing.
The control of traveling quantum fields encoded with information (flying qubit)  is crucial for quantum networks.
In this work, we propose to catch and release the propagating photon/phonon with a non-Markovian giant atom, which is coupled to the environment via multiple coupling points.
Based on the Heisenberg equation of motion for the giant atom and field operators, we calculate the time-dependent  scattering coefficients from the linear response theory and define the criteria for the non-Markovian giant atom.
We analyze and numerically verify that the field bound states due to non-Markovianity can be harnessed to catch and release the propagating bosonic field on demand by tuning the parameters of giant atom.
\end{abstract}

\title[Catch and release of propagating bosonic field with non-Markovian giant atom]
\maketitle

\section{Introduction}

In standard quantum optics, the (natural or artificial) atom is usually treated as a point-like model that is locally coupled to bosonic modes in the environment, where the dipole approximation is well justified \cite{Walls2008}. In recent years, however, the artificial atom realized in the laboratory can be  made much larger than the wavelength of bosonic modes in the environment, and thus is dubbed \textit{giant atoms} \cite{anton2021book}. Such  giant-atom model has been implemented with superconducting transmon \cite{Koch2007pra}  coupled to surface acoustic waves (SAWs) \cite{Gustafsson2014science,Andersson2019np} or microwave photons \cite{Kannan2019,vadiraj2021pra} in a nonlocal fashion via multiple coupling points. The experimental platform has also been extended to magnon-based hybrid system, where a ferro-magnetic spin ensemble interacts twice with the meandering waveguide \cite{Wang2022nc}.
Giant atoms distinguish from small atoms on two main aspects: (1) the self-interference effect induced by the system size \cite{Kockum2014pra}; and (2) the non-Markovian effect due to the time delays among the coupling points \cite{Guo2017}.
%
The self-interference of giant atom gives rise to many novel phenomena that do not exist for small atoms, e.g.,  frequency-dependent relaxation rates and Lamb shifts \cite{Kockum2014pra}, decoherence-free interaction between giant atoms \cite{Kockum2018prl,carollo2020prr}.
For the non-Markovian giant atom, more intriguing phenomena can appear, such as non-exponential spontaneous emission \cite{Guo2017,Andersson2019np} and persistent oscillating bound states \cite{guo2020prr,taylor2020pra}.

The study of giant atoms with nonlocal light-matter interaction represents a new paradigm and research frontier in quantum optics \cite{Kockum2019b,Gonzalez-Tudela2019,Calajo2019,Lorenzo2021sr}.
The  original theoretical framework \cite{Kockum2014pra,Guo2017} of giant atom(s) coupled to waveguide with linear dispersion relationship \cite{sheremet2023rmp} has been extended to that of giant atom(s) interacting with structured environment, e.g., coupled resonators/cavities \cite{zhao2020pra,lim2023pra,zhang2023pra,jia2023atomphoton,bag2023quantum} or linear waveguide with designed spatial coupling sequence \cite{wang2023realizing}.
The giant-atom effects in optical regime is also proposed with two coupled Rydberg atoms \cite{chen2023giantatom}. Giant atom in two dimensional environment was proposed with atomic matter wave in optical lattice \cite{Gonzalez-Tudela2019}.
The interplay between giant atoms and topological environment has also drawn more and more attention  \cite{cheng2022pra,vega2021pra,vega2023prr}. Recent work unveils the behavior of giant atoms coupled to a non-Hermitian Hatano-Nelson (HN) mode due to the asymmetric inter-site tunneling rates \cite{du2023giant}.
Giant atoms with chiral quantum optics \cite{wang2023realizing,Guimond2020,joshi2023prx,zhang2021prx,Kannan2023nat,Wang2022qst,du2022prl,vega2023prr} have opened new possibilities to realize chiral or nonreciprocal phenomena such as directed emission  and chiral bound states  \cite{li2023tunable,wang2021prl}.
The entanglement dynamics  \cite{yin2023pra}  and non-Markovian disentanglement dynamics \cite{yin2022pra} in double-giant-atom waveguide-QED systems were studied  for separated, braided, and nested coupling configurations. Maximally entangled long-lived states  of giant atoms can be generated with  photon-mediated interactions in a waveguide when the system is driven by a resonant classical field  \cite{santos2023prl}.
The two-level giant atom has been extended to three-level artificial atom \cite{zhu2022pra,Chen2022cp,gu2023correlated}. A recent trend of giant atom study is to explore the physics in ultrastrong coupling regime \cite{Ask2019,noachtar2022pra,zueco2022pra} and multi-photon/phonon process \cite{Guo2017,Guimond2017,gu2023correlated,cheng2023lp,arranz2021prr,pichler2016prl}. Last but not least, the real-space environment that interacts with giant atoms  can be extended to a synthetic frequency dimension \cite{du2022prl}.

Giant atoms not only represent another paradigm change in quantum optics but also can be exploited for designing functional quantum devices, e.g.,
unidirectional microwave photon emitter (absorber) \cite{Guimond2020,Kannan2023nat}  and single photon/phonon router \cite{Chen2022cp,Wang2021oe}, that are suitable for building interconnects within extensible quantum networks. Note that the capabilities of these functional quantum devices are mostly realized based on the size-induced self-interference effects of giant atom \cite{Guo2017,zhu2022pra,Qiu2023,yin2022pra,du2023pra,lim2023pra,cheng2022pra,ask2022prl,arranz2021prr,pichler2016prl,yin2022pra}. In fact, non-Markovianity due to the time delays among the legs of giant atoms  can also be  a valuable resource for quantum information processing \cite{hannes2017pans,Sletten2019}. In this work, we propose that a two-leg non-Markovian giant atom can be utilized to tweeze (catch and release) the propagating field in the environment on demand. Such \textit{photon/phonon tweezer} may find its application as a hardware-efficient memory unit for quantum signal processing \cite{hann2019prl} in quantum networks, where the precise control of traveling field is highly demanding for large-scale networked quantum systems \cite{divincenzo1995sci}.

Our paper is organized as follows. In Section~\ref{sec-model}, we introduce the experimental setups and the model Hamiltonian of giant atom with multiple coupling points. In Section~\ref{sec-EOM}, we derive the Heisenberg equation of motions (EOM) for the operators of giant atom and bosonic field in the harmonic limit, which are then analytically solved with Laplace transformation method. In Section~\ref{sec-scattering}, we investigate the scattering dynamic and calculate the scattering coefficients based on the linear response theory. From the scattering spectrum, we obtain a criteria to define the non-Markovian regime for a giant atom. In Section~\ref{sec-tweezer}, we investigate and numerically verify the proposal to catch and release propagating bosonic field with a two-leg non-Markovian giant atom, which can be implemented by a circuit-QED architecture.  In Section~\ref{sec-sumlook}, we summarize the results in the paper and outlook the research directions in the future to design more non-Markovian quantum devices with giant atoms.

\section{Experimental  Setup and Model Hamiltonian}\label{sec-model}

\begin{figure}
  \centering
  \includegraphics[scale=0.5]{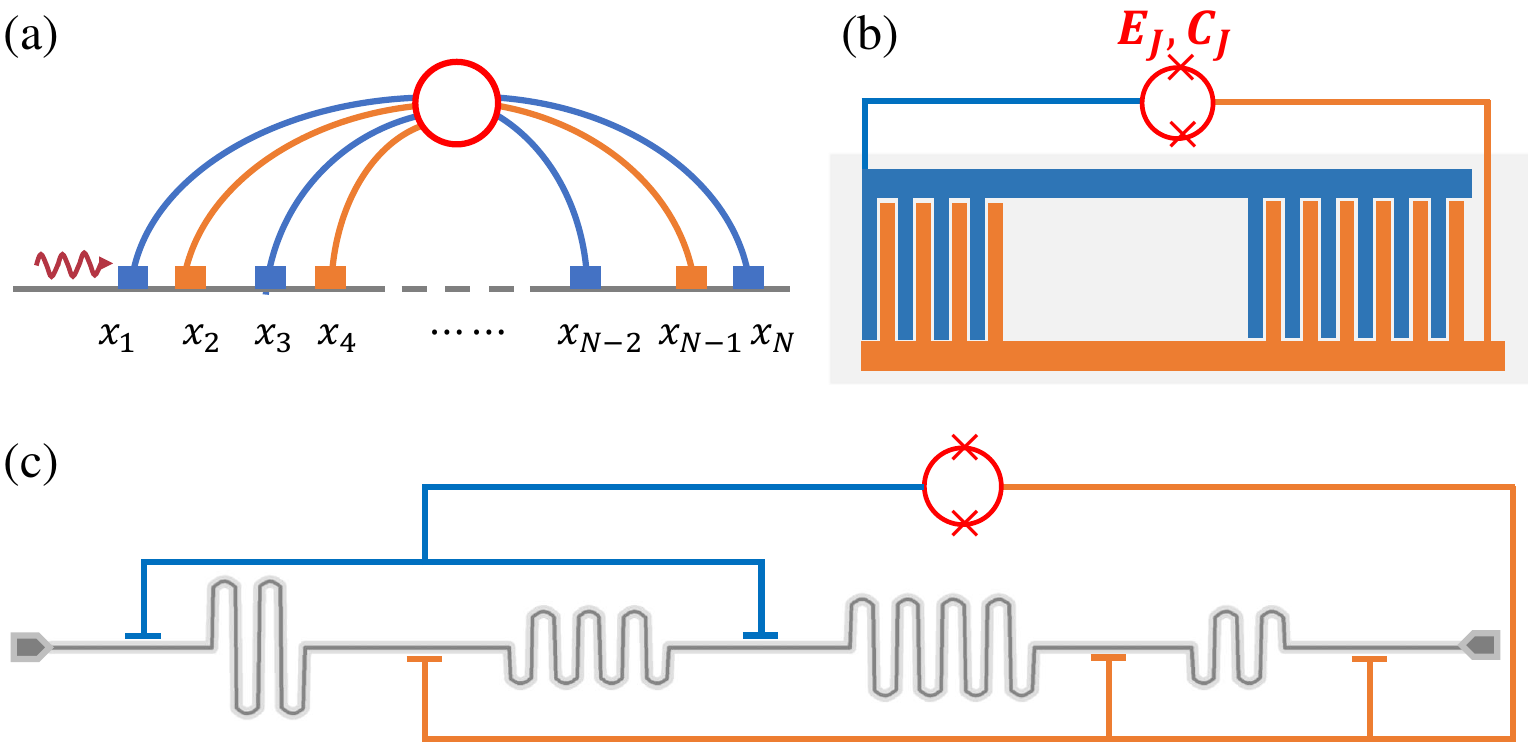}\\
  \caption{{\bf{Devices:} Sketch and experimental setups of giant atom.}
  {\bf (a)} Sketch of theoretical model: an atom (empty circle) coupled to a transmission line (gray) at multiple points $x_j$ with positive (blue)  and negative (orange) coupling strength.
  {\bf (b)} Experimental schematic of a transmon  capacitively coupled to surface acoustic waves (SAWs) via interdigital transducer (IDT, orange and blue finger-like structures).
  The giant atom is implemented by the transmon that is made up of two Josephson junctions forming a SQUID loop with tunable Josephson energy $E_J$ and capacitance $C_J$.
  The SAWs propagating on the substrate (gray) are coupled to transmon via piezoelectric mechanism.
 {\bf (c)} Experimental schematic of a transmon coupled capacitively to a microwave transmission line via multiple points. The distance between coupling points can be tuned by meandering the transmission line.
   }\label{fig setup}
\end{figure}

The experimental device we consider is made up of an artificial  atom (transmon) \cite{Barends2013,Koch2007pra} coupled to, e.g., the piezoelectric waveguide via designed metallic fingers as shown in Fig.~\ref{fig setup}(b), or the one-dimensional (1D) meandering transmission line via multiple coupling points as shown in Fig.~\ref{fig setup}(c). The bosonic modes in the piezoelectric waveguide and the 1D transmission line are the surface acoustic waves (SAWs)  \cite{Rayleigh1885PLMS,Datta1986SAW,white2004book}  and electromagnetic microwaves respectively. In both experimental setups, the artificial giant atom is coupled to  the \emph{electric potential field} $\phi(x,t)$ that is induced either by SAWs via piezoelectric mechanism or confined directly in the 1D transmission line.
The theoretical model is sketched by Fig.~\ref{fig setup}(a) and the total system Hamiltonian can be split into three parts $H_s=H_{tr}+H_{tL}+H_{int}$, where $H_{tr}$,  $H_{tL}$ and $H_{int}$ represent the transmon (atom) Hamiltonian, the field Hamiltonian and their interaction Hamiltonian respectively.

%
%

%
\subsection{Artificial atom}

The transmon, which plays the role of artificial atom in our model, is made up of a SQUID loop with Josephson energy $E_J$ and coupled capacitively to the bosonic field modes in the transmission line (waveguide).
The Hamiltonian of transmon including its interaction with bosonic field is described by \cite{Koch2007pra}
\begin{eqnarray}\label{Transmon}
H_{tr}+H_{int}&=&\frac{(2e)^2}{2C_\Sigma}(\hat{n}-\hat{n}_s)^2-E_J\cos\hat{\eta}\nonumber\\
&\equiv&4E_C\hat{n}^2-E_J\cos\hat{\eta}-8E_C\hat{n}\hat{n}_s+4E_C\hat{n}_s^2.
\end{eqnarray}
Here, $\hat{n}_s$ is the offset charge of the transmon measured in units of the Cooper pair charge $2e$,  and $E_C={e^2}/{2C_\Sigma} $ is the charging energy of transmon with $C_\Sigma$ the total capacitance. The two conjugate observables of transmon are the net number of Copper pairs $\hat{n}$ and the phase difference $\hat{\eta}$ across the Joesphson junctions (JJs) of SQUID.
 By defining the ladder operator $\hat{b}\equiv\big(\frac{E_J}{32E_C}\big)^{\frac{1}{4}}\hat{\eta}+i\big(\frac{2E_C}{E_J}\big)^{\frac{1}{4}}\hat{n}$, the free transmon Hamiltonian is
\begin{eqnarray}\label{Transmon2}
H_{tr}=4E_C\hat{n}^2-E_J\cos\hat{\eta}\approx\hbar\Omega(\hat{b}^\dag\hat{b}+\frac{1}{2})-\kappa(\hat{b}+\hat{b}^\dag)^4,
\end{eqnarray}
where $\Omega=\sqrt{8E_CE_J}/\hbar$ is known as the Josephson plasma frequency
and
$\kappa={E_C}/{12}$
is the nonlinearity to the 4th-order expansion of Josephson potential.

\subsection{Bosonic field}

The bosonic mode that directly couples to the transmon is the electrical potential field $\phi(x,t)$. Its corresponding conjugate field
is the flux field $\Phi(x,t)\equiv\int^tdt'\phi(x,t')$ satisfying
classical wave equations
$(\partial_t^2-v^2\partial_x^2)\Phi(x,t)=0$
with $v$ the velocity of propagating field. Assuming a linear dispersion relationship of bosonic modes $\omega_k=|k|v$ with $k$ the wavenumebr, the quantized flux field and
 electric potential field are \cite{Peropadre2013}
\begin{eqnarray}
\hat{\Phi}(x,t)&=&\sqrt{\frac{\hbar Z_0v}{4\pi}}\int_{-\infty}^\infty
dk\frac{1}{\sqrt{\omega_k}}\Big(\hat{a}_k e^{-i(\omega_k t-
kx)}+{\rm H.c.}\Big),\label{Phi-1}\\
\hat{\phi}(x,t)
&=&-i\sqrt{\frac{\hbar Z_0v}{4\pi}}\int_{-\infty}^\infty
dk\sqrt{\omega_k}\Big(\hat{a}_k e^{-i(\omega_k t- kx)}-{\rm H.c.}\Big).\label{Phi-2}
\end{eqnarray}
Here, $Z_0$ is the characteristic impedance of transmission line, $\hat{a}_k$ is the annihilation operator of the bosonic mode with wavenumber $k$ satisfying $[\hat{a}_k,\hat{a}^\dagger_{k'}]=\delta(k-k')$. From Eqs.~(\ref{Phi-1}) and (\ref{Phi-2}) , we have
$[\hat{\Phi}(x,t),\hat{\phi}(x',t)]=i\hbar Z_0v\delta(x-x')$. The quantized Hamiltonian of bosonic field modes is
\begin{eqnarray}\label{HTL}
H_{tL}=\int_{-\infty}^{+\infty}\  dk\  \hbar\omega_k\
\hat{a}_k^\dag\hat{a}_{k}.
\end{eqnarray}

\subsection{Interaction}
The coupling between transmon and bosonic field mode is
described by the $H_{int}=-8E_C\hat{n}\hat{n}_s$ in
Eq.(\ref{Transmon}). The offset charge  is $\hat{n}_s
=\frac{1}{2e}\sum_m C_{m}\hat{\phi}(x_{m},t)$,
where $C_{m}$ represents the effective capacitance of each coupling point at the position $x_m$.
%
%
From Eq.~(\ref{Phi-2}), we obtain the interaction Hamiltonian
\begin{eqnarray}\label{Int}
H_{int}
&=&\frac{4E_C}{e}\sqrt{\frac{1}{2\eta}}\sqrt{\frac{\hbar
Z_0v}{4\pi}}\sum^N_{m=1}C_{m}\int_{-\infty}^\infty
dk\sqrt{\omega_k}\Big(\hat{a}_k
e^{ikx_{m}}-{\rm H.c.}\Big)(\hat{b}-\hat{b}^\dag)\nonumber\\
&\approx&-\frac{4E_C}{e}\sqrt{\frac{1}{2\eta}}\sqrt{\frac{\hbar
Z_0v}{4\pi}}\sum^N_{m=1}C_{m}\int_{-\infty}^\infty
dk\sqrt{\omega_k}\Big( \hat{b}^\dag\hat{a}_k
e^{ikx_{m}}+{\rm H.c.}\Big).
\end{eqnarray}
%
%
In the second line, we have  adopted the rotating wave approximation (RWA) by dropping off the counter-rotating terms like $\hat{a}^\dagger_k\hat{b}^\dagger$ and $\hat{a}_k\hat{b}$.
By introducing the coupling strength at each coupling point
$c_m\equiv-\frac{4E_C}{e}\sqrt{\frac{1}{2\eta}}\sqrt{\frac{\hbar
Z_0v}{4\pi}}C_{m},$
 the total system Hamiltonian is simplified as
\begin{eqnarray}\label{TotalH}
H_s&=&\hbar\Omega(\hat{b}^\dag\hat{b}+\frac{1}{2})-\kappa(\hat{b}+\hat{b}^\dag)^4
+\int_{-\infty}^{+\infty}\  dk\  \hbar\omega_k\
\hat{a}_k^\dag\hat{a}_{k}\nonumber\\
&&+\sum_{m=1}^{N}c_{m}\int_{-\infty}^{+\infty}
\Big(e^{ikx_{m}}\hat{a}_k\hat{b}^\dag+{\rm H.c.}\Big)\sqrt{\omega_k}dk.
\end{eqnarray}
Here in our work, we allow the coupling legs of giant atom to stretch from both sides of transmon, cf., blue and orange colors in Fig.~\ref{fig setup}. As a result, the coupling strength $c_m$ belonging to different groups of coupling legs can be either positive or negative.


\section{Heisenberg equation of motion and solution}\label{sec-EOM}

Given the system Hamiltonian, we now derive the equation of motion (EOM) for the giant atom operator $\hat{b}$ and the field operator $\hat{a}_k$.
In this work, we set the transmon Hamiltonian in the harmonic limit [$\kappa \to 0$ in Eq.~(\ref{TotalH})]. In this case, the system keeps in the coherent states during time evolution if we prepare all the hamonic modes in coherent states initially. As a result, the EOM of the averaged atomic operator  over the coherent state follows the same EOM for the atomic probability amplitude in the single-excitation approximation \cite{guo2020prr}. Therefore, all the conclusions obtained in this work are valid either for the multi-level atom without nonlinearity  (a harmonic oscillator) or the two-level atom in the single photon/phonon process.

\subsection{Heisenberg equation of motion}
The Heisenberg
EOMs for the atomic operator $\hat{b}(t)$ and the field operator $\hat{a}_k(t)$ are given by
\begin{eqnarray}\label{EOMb}
\frac{d}{dt}\hat{b}(t)&=&\frac{1}{i\hbar}[\hat{b}(t),H_{s}]
=-i\Omega\hat{b}(t)-\frac{i}{\hbar}\sum_{m}c_{m}\int_{-\infty}^\infty
e^{ikx_{m}}\hat{a}_k(t) \sqrt{\omega_k}dk\label{EOMb-1}\\
\frac{d}{dt}\hat{a}_k(t)&=& -\frac{i}{\hbar}[\hat{a}_k(t),H_{s}]=-i\omega_k \hat{a}_k(t)-i\frac{\sqrt{\omega_k}}{\hbar}\sum_mc_me^{-ikx_m}\hat{b}(t).\label{EOMb-2}
\end{eqnarray}
Formally integrating the EOM (\ref{EOMb-2}) of the field operator, we have
\begin{eqnarray}\label{akt}
\hat{a}_k(t)=e^{-i\omega_k t}\left[\hat{a}_k(0)-i\frac{\sqrt{\omega_k}}{\hbar}\sum_mc_me^{-ikx_m}\int_0^tdt'e^{i\omega_kt'}\hat{b}(t')\right].
\end{eqnarray}
Inserting Eq.~(\ref{akt}) into Eq.~(\ref{EOMb}), we obtain
\begin{eqnarray}\label{bt0}
\frac{d}{dt}\hat{b}(t)
&=&-i\Omega\hat{b}(t)
-\frac{i}{\hbar}\sum_{m}c_{m}\int_{-\infty}^\infty \sqrt{\omega_k}
e^{ikx_{m}}e^{-i\omega_kt }\hat{a}_k(0) dk\nonumber\\
&&-\frac{1}{\hbar^2v}\sum_{m,m'}c_{m}c_{m'}\int_0^tdt'\hat{b}(t')\int_{0}^\infty
e^{i\omega_k(|\tau_{mm'}|+t'-t)} \omega_kd\omega_k.
\end{eqnarray}
Here, we have used the linear dispersion relation of field modes $\omega_k=v|k|$, and defined the delay time $\tau_{mm'}\equiv (x_m-x_{m'})/v$ between two coupling points at $x_m$ and $x_{m'}$.

In order to further simplify Eq.~(\ref{bt0}), we adopt the well-known Weisskopf-Wigner approximation\cite{Scully1997} for each coupling point. Since the intensity of the emitted radiation is dominant in the range around the atomic transition frequency $\Omega$, we can approximate the integral over the frequency around  $\omega_k\approx\Omega$ and replace $d\omega_k$ by $d(\omega_k-\Omega)$ in the last term of Eq.~(\ref{bt0}).
By setting the lower limit as negative infinity ($-\infty$) and using the identity $\int_{-\infty}^{+\infty}e^{ixt}dt=2\pi\delta(x)$, the  Eq.~(\ref{bt0}) is simplified as
\begin{eqnarray}\label{Solution of EOM1}
\frac{d}{dt}\hat{b}(t)
&\approx&-i\Omega\hat{b}(t)
-i\sqrt{\frac{\Gamma v}{2\pi}}\sum_{m}c_{m}\int_{-\infty}^\infty
e^{ikx_{m}}e^{-i\omega_kt }\hat{a}_k(0) dk\nonumber\\
&&-\Gamma\sum_{m,m'}c_mc_{m'}\int_0^t \delta(t-t'-|\tau_{mm'}|)\hat{b}(t')dt'\nonumber\\
&=&-i\Omega\hat{b}(t)
-i\sqrt{\frac{\Gamma v}{2\pi}}\sum_{m}c_{m}\int_{-\infty}^\infty
e^{ikx_{m}-i\omega_kt}\hat{a}_k(0) dk\nonumber\\
&&
-\Gamma
\sum_{m,m'}c_mc_{m'}\hat{b}(t-|\tau_{mm'}|)\Theta(t-|\tau_{mm'}|).
\end{eqnarray}
Here, we have introduced the parameter
$\Gamma\equiv\frac{2\pi\Omega}{\hbar^2
 v}$,
and
$\Theta(x)$ is the Heaviside step function.
Since the EOM for $\hat{b}(t)$ is linear, we assume the solution to be the following form
\begin{eqnarray}\label{Solution of EOM2}
 \hat{b}(t)&\equiv& \chi(t)\hat{b}(0)+\int_{-\infty}^\infty dk
\xi_k(t)\hat{a}_k(0)
\end{eqnarray}
with initial conditions $\chi(0)=1$ and $\xi_k(0)=0$.
Plugging this solution into Eq.~(\ref{Solution of
EOM1}), we have
\begin{eqnarray}\label{Solution of EOM3}
\frac{d}{dt}\chi(t)&=&-i\Omega\chi(t)-\Gamma\int_0^tg(t-t')\chi(t')dt',\label{Solution of EOM3-1}\\
\frac{d}{dt}\xi_k(t)&=&-i\Omega\xi_k(t)-\Gamma\int_0^tg(t-t')\xi_k(t')dt'-i\sqrt{\frac{\Gamma v}{2\pi}}\sum_mc_m
 e^{ikx_m-i\omega_kt},\label{Solution of EOM3-2}
\end{eqnarray}
where we have defined the memory function
\bea
g(t-t')=\sum_{m,m'}c_mc_{m'}\delta(t-t'-|\tau_{mm'}|).
\eea
 Since both the atom operator $\hat{b}(t)$ and field operator $\hat{a}_k(0)$ have to satisfy
the noncommutative relationship $[\hat{b}(t),\hat{b}^\dagger(t)]=1$ and $[\hat{a}_k(0),\hat{a}^\dagger_{k'}(0)]=\delta(k-k')$, we have the restriction condition $|\chi(t)|^2+\int_{-\infty}^\infty dk
|\xi_k(t)|^2=1$.

In the harmonic limit, the Hamiltonian~(\ref{TotalH}) describes the linear problem where a harmonic oscillator interacts with the continuum of harmoinc modes in the transmission line.
If the system is initially prepared in the coherent states with coherent values $\langle \hat{b}(0)\rangle=\beta(0)$ and $\langle \hat{a}_k(0)\rangle=\xi_k(0)$ for the atom state and filed mode respectively, the atom will keep staying in the coherent state $|\beta(t)\rangle$ with coherent value given by
\bea
\beta(t)\equiv\langle \hat{b}(t)\rangle=\chi(t)\beta(0)+\int_{-\infty}^\infty dk \xi_k(t)\alpha_k(0).
\eea
As already proved by Ref.~\cite{guo2020prr}, the complex harmonic amplitude $\beta(t)=\langle \hat{b}(t)\rangle$ follows exactly the same EOM as that of the probability amplitude for the two-level atom in the single-excitation approximation.
%

In order to solve the EOMs of  $\chi(t)$ and $\xi_k(t)$ given by Eqs.~(\ref{Solution of EOM3-1}) and (\ref{Solution of EOM3-2}), we perform the Laplace transformation $E_f(s)\equiv\int^{\infty}_0 dt f(t)e^{-s t}$ with $f$ representing $\chi(t)$ or $\xi_k(t)$, and obtain
%
%
\begin{eqnarray}
E_\chi(s)&=&\frac{1}{s+i\Omega+\Gamma\sum_{m,m'}c_mc_{m'}e^{-s|\tau_{mm'}|}},\label{Echi}\\
E_{\xi_k}(s)&=&-i\sqrt{\frac{\Gamma v}{\pi}} \frac{\sum_m
 c_me^{ikx_m}}{(s+i\omega_k)[s+i\Omega+\Gamma\sum_{m,m'}c_mc_{m'}e^{-s|\tau_{mm'}|}]}.\label{Eb}
\end{eqnarray}
The time evolution of $\chi(t)$ and $\xi_k(t)$ can be obtained by applying the inverse Laplace transformation to Eqs.~(\ref{Echi}) and
(\ref{Eb})
\begin{eqnarray}\label{InverseEs}
\chi(t)&=&\sum e^{s_{ pole}t}\mathrm{Res}(E_\chi(s),s_{ pole})
=\sum e^{s_{ pole}t}\lim_{s\rightarrow
s_{ pole}}E_\chi(s)(s-s_{ pole}),\nonumber\\
\xi_k(t)&=&\sum e^{s_{ pole}t}\mathrm{Res}(E_{\xi_k}(s),s_{ pole})
=\sum e^{s_{ pole}t}\lim_{s\rightarrow
s_{ pole}}E_{\xi_k}(s)(s-s_{ pole}),
\end{eqnarray}
where $\mathrm{Res}(E_f(s),s_{pole})$ is the residue of $E_f(s)$ at the pole $s_{pole}$. From Eqs.~(\ref{Echi}) and
(\ref{Eb}),
the poles of  $E_\chi(s)$ and $E_{\xi_k}(s)$ are $s_{pole}=-i\omega_k $, and also
given by the roots (labeled $s_n$) of the following equation
\begin{eqnarray}\label{polesEq}
s+i\Omega+\Gamma\sum_{m,m'}c_mc_{m'}e^{-s|\tau_{mm'}|}=0.
\end{eqnarray}
Thus, for $t>0$, we have the solutions as following:
\begin{eqnarray}
 \chi(t)&=&\sum_{n}\frac{e^{s_n
t}}{1-\Gamma\sum_{m,m'}c_mc_{m'}|\tau_{mm'}|e^{-s_n|\tau_{mm'}|}},\label{PolesI}\\
\xi_k(t)&=&-i\sqrt{\frac{\Gamma v}{2\pi}}\sum_n\
 \frac{\sum_{m}c_m
 e^{ikx_m}e^{s_nt}}{(s_n+i\omega_k)[1-\Gamma\sum_{m,m'}c_mc_{m'}|\tau_{mm'}|e^{-s_n|\tau_{mm'}|}]}\nonumber\\
&&
-i\sqrt{\frac{\Gamma v}{2\pi}}\
 \frac{\sum_mc_m
 e^{ikx_m}e^{-i\omega_kt}}{i(\Omega-\omega_k)+\Gamma\sum_{m,m'}c_mc_{m'}e^{i\omega_k|\tau_{mm'}|}}.\label{PolesII}
\end{eqnarray}
If the pole $s_{pole}=-i\omega_k$ also satisfies Eq.~(\ref{polesEq}), the transmon is decoupled from the mode $\omega_k=i s_n$, and the value of $\xi_k(t)$ is zero.
Given the solution of $\hat{b}(t)$, we obtain the field operator $\hat{\varphi}(x,t)$ at position $x$ and time $t$ in the transmission line from Eq.(\ref{akt})
\begin{eqnarray}\label{numberdensity}
\hat{\varphi}(x,t)
&\equiv&\frac{1}{\sqrt{2\pi}}\int_{-\infty}^\infty dk
 e^{ikx}\ \hat{a}_k(t)\nonumber\\
&\approx&-i\sqrt{\frac{\Gamma }{v}} \sum_{m}c_{m}\hat{b}(t-\frac{|x-x_m|}{v})\Theta(t-\frac{|x-x_m|}{v})\nonumber\\
 &&+\frac{1}{\sqrt{2\pi}}\int_{-\infty}^\infty dk
 e^{i(kx-\omega_kt)}\ \hat{a}_k(0).
\end{eqnarray}
Here, we used the Weisskopf-Wingner approximation again as we did in Eq.~(\ref{bt0}). The density of field in the transmission line is given by $p(x,t)=|\langle  \hat{\varphi}(x,t)\rangle|^2$, which also describes the probability density to find a photon/phonon with all the possible wavenumber $k$.

\subsection{Effective model}
%
%

As shown in Fig.~\ref{fig setup}(b), in the experiment of giant atom coupled to the SAW field \cite{Gustafsson2014science,Sletten2019,Andersson2019np}, several separated IDTs  are often arranged coupled to the same transmission line with long distance (e.g., hundreds of SAW wavelengths), while the distance of neighboring fingers inside each IDT is close to SAW wavelength. In this section, we prove that the separated IDTs far apart from each other can be simplified to an effective model,  in which the each IDT can be treat integrally as a single coupling point with the effective coupling strength that can be tuned by the system parameters.

\begin{figure}
  \centering
  \includegraphics[scale=0.4]{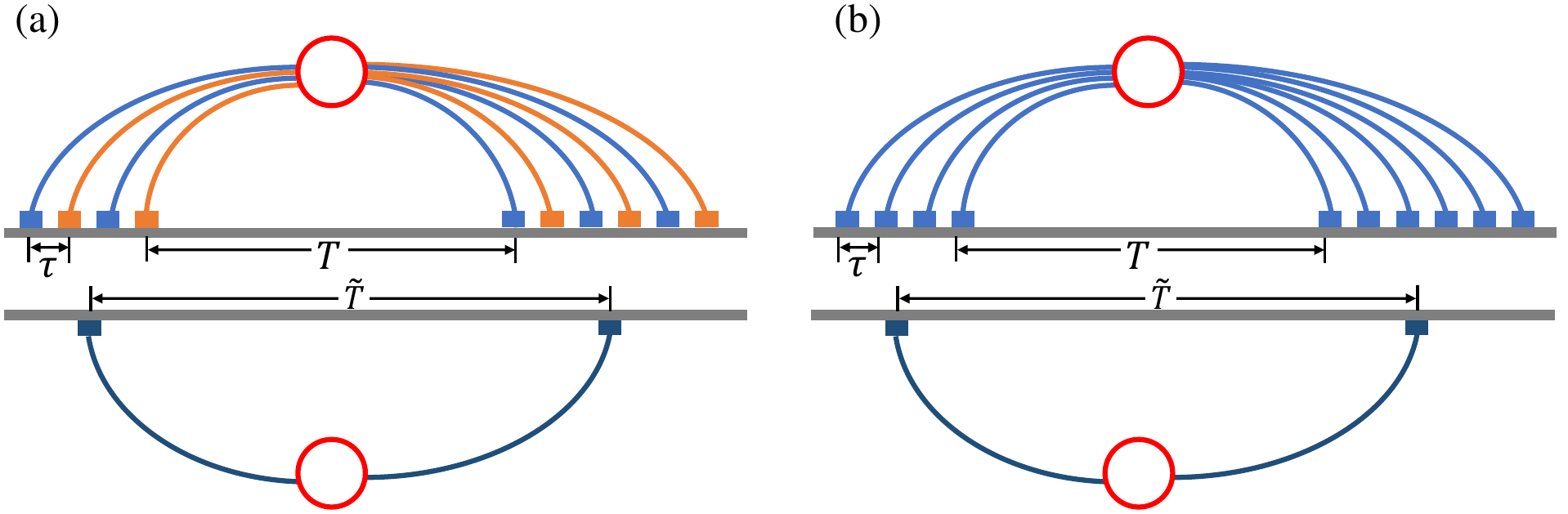}\\
  \caption{{\bf{Effective models:}} a giant atom (empty circle) with multiple legs that are divided into two separate groups.   The signs of coupling strength of the giant atom's legs  in each group is set to be  {\bf{(a)}} positive and negative alternately; {\bf (b)} positive uniformly. The lower halves in both figures show the simplified effective giant-atom model with two effective coupling points that locate at the center of each group. Parameters $\tau$, $T$ and $\tilde{T}$ represent the delay times of bosonic field propagating between the neighboring legs in each group, the two groups and the two effective coupling points, respectively.
 }\label{fig 2IDT}
\end{figure}

As shown in Fig.~\ref{fig 2IDT}, we consider a giant atom model in which the total coupling points can be divided into two separated parts that locate far way from each other. 
Without loss of generality, we assume the coupling legs are uniformly distributed in each part and the numbers of legs in the left and right parts are even integers $N_1$ and $N_2$, respectively, with $N_1\leq N_2$.
In Fig.~\ref{fig 2IDT}(a), we set the single-leg coupling strength to be $c_{m_1}=(-1)^{m_1}c_1$  with  $m_1=1,2,\cdots,N_1$ and
$c_{m_2}=(-1)^{m_2}c_2$ with $m_2=N_1+1,\cdots, N_1+N_2$. In Fig.~\ref{fig 2IDT}(b), we set  the single-leg coupling strength to be uniform, i.e., $c_{m_1}=c_1$ with $m_1=1,2,\cdots,N_1$ and
$c_{m_2}=c_2$ with $m_2=N_1+1,\cdots, N_1+N_2$.
We set $\tau$ as the delay time between the neighboring legs in the same part, and $T$ as the delay time between the rightmost leg of the left part and the leftmost leg of right part. 
The delay times are obtained by combining arbitrary two coupling points. As a result, all the possible delay times $|\tau_{mm'}|$ can be classified into two groups: 1) the delay times traveling inside each part, i.e., $|\tau_{mm'}|=n_j\tau$ with
$n_j=0,1,2,\cdot\cdot\cdot, N_j-1$, where $j=1,2$ labels the two separated parts; 2) the delay times across the two separated parts, i.e., $|\tau_{mm'}|=T+n\tau$ with $n=0,1,2,\cdot\cdot\cdot, N_1+N_2-2$. For the delay times of first group, the combination number is $N_j$ for $n_j=0$; and $2(N_j-n_j)$ for $n_j>0$. For the delay times of second group, the combination number is $2(n+1)$ for $n < N_1$; $2N_1$ for $N_1\leq n<N_2$; and $2(N_1+N_2-n-1)$ for $n\geq N_2$, respectively.

For the case of Fig.~\ref{fig 2IDT}(a), assuming the bosonic field is in the ground state $\langle\hat{a}_k(0)\rangle=0$,  we have EOM of
$\beta(t)\equiv\langle\hat{b}(t)\rangle$ from Eq.~(\ref{Solution of EOM1})
\begin{eqnarray}\label{EOM2IDT}
\frac{d}{dt}\beta(t)&=&-i\Omega \beta(t)-\Gamma
\sum_{m,m'}c_m c_{m'} \beta(t-|\tau_{mm'}|)\Theta(t-|\tau_{mm'}|)\nonumber\\
&=&(-i\Omega +N_1\gamma_1/2
+N_2\gamma
_2/2)\beta(t)\nonumber\\
&&-\sum_{j=1,2}
\gamma_j\sum_{n_j=0}^{N_j-1}(-1)^{n_j}(N_j-n_j)\beta(t-n_j\tau)\Theta(t-n_j\tau)\nonumber\\
&&-\sqrt{\gamma_1\gamma_2}\left[\sum_{n=0}^{N_1-1}(n+1)+\sum_{n=N_1}^{N_2-1}N_1
+\sum_{n=N_2}^{N_1+N_2-2}(N_1+N_2-n-1)\right]\nonumber\\
&&\times(-1)^{n+1}\beta(t-T-n\tau)\Theta(t-T-n\tau).
\end{eqnarray}
Here, we have defined parameters $\gamma_1=2\Gamma c_1^2$, and $\gamma_2=2\Gamma c_2^2$.
Note that the above EOM is written in the lab frame. In the RWA, the dynamics of $\beta(t)$ can be separated into a fast oscillation with frequency $\Omega$ and a low time evolution of amplitude $\beta(t)=\tilde{\beta}(t)e^{-i\Omega t }$.
If the longest delay times inside the two separated parts $N_1\tau$, $N_2\tau$ are much shorter than the characteristic time of $\tilde\beta(t)$, we are allowed to neglect the delay time differences $n\tau$ in $\tilde{\beta}(t-n\tau)$, and replace the delay time differences $T+n\tau$ in $\tilde{\beta}(t-T-n\tau)$ by the traveling time between the centers of two separated parts $\tilde{T}=T+(N_1+N_2)\tau/2-\tau$, cf. the lower subfigure of Fig.~\ref{fig 2IDT}(a).
In the end, the EOM of $\beta(t)$ is equivalent  into the simplified model with two effective coupling legs (see detailed derivation in ~\ref{AppA})
\begin{eqnarray}\label{EOM2IDT-1}
\frac{d}{dt}\beta(t)&=&-i(\Omega+\tilde{\Delta})\beta(t)
-\tilde{\gamma}_1\beta(t)-\tilde{\gamma}_2\beta(t-\tilde{T})\Theta(t-\tilde{T}),
\end{eqnarray}
with the Lamb shift $\tilde{\Delta}$ and the effective decay rate $\tilde{\gamma}_1, \tilde{\gamma}_2$ defined as
\begin{eqnarray}
\tilde{\Delta}&=&-\frac{1}{2} \sum_{j=1,2}\gamma_j\frac{N_j\sin (\Omega \tau)+\sin(N_j\Omega\tau)}{1+\cos\Omega\tau},\label{EOM2IDT-2a}\\
\tilde{\gamma}_1&=&\frac{1}{2} \sum_{j=1,2}\gamma_j\frac{1-\cos(N_j\Omega\tau)}{1+\cos\Omega\tau},\label{EOM2IDT-2b}\\
\tilde{\gamma}_2&=&\sqrt{\gamma_1\gamma_2}\frac{\cos(\Delta N\Omega\tau)-\cos(\bar{N}\Omega\tau)}{1+\cos\Omega\tau}.\label{EOM2IDT-2c}
\end{eqnarray}
Here, $\bar{N}=(N_1+N_2)/2$ and $\Delta N=(N_2-N_1)/2$.
%

For another case shown in Fig.~\ref{fig 2IDT}(b), the EOM of $\beta(t)$ can also be expressed by the simplified model of Eq.~(\ref{EOM2IDT-1}) but with Lamb shift $\tilde{\Delta}$ and effective decay rates $\tilde{\gamma}_1, \tilde{\gamma}_2$ given by (see detailed derivation in ~\ref{AppA})
 \begin{eqnarray}
\tilde{\Delta}&=&\frac{1}{2}\sum_{j=1,2}\gamma_j\frac{N_j\sin (\Omega \tau)-\sin(N_j\Omega\tau)}{1-\cos\Omega\tau}\label{EOM2IDT-3a}\\
\tilde{\gamma}_1&=&\frac{1}{2} \sum_{j=1,2}\gamma_j\frac{1-\cos(N_j\Omega\tau)}{1-\cos\Omega\tau}\label{EOM2IDT-3b}\\
\tilde{\gamma}_2&=&\sqrt{\gamma_1\gamma_2}\frac{\cos(\Delta N\Omega\tau)-\cos(\bar{N}\Omega\tau)}{1-\cos\Omega\tau}.\label{EOM2IDT-3c}
\end{eqnarray}
%
Both the Lamb shift and the strength of the coupling points depend on the atom frequency $\Omega$, the finger numbers, accumulating phase $\Omega\tau$ and  coupling strength distributions.
 Note that, for the experimental setup as shown in Fig.~\ref{fig setup}(b), if we set each finger pair as a single leg, Eqs.~(\ref{EOM2IDT-3a})-(\ref{EOM2IDT-3c}) are also applied as long as we double the delay time $\tau$ and replace the capacitance of each leg by that of a finger pair \cite{Andersson2019np}.

\section{Scattering dynamics}\label{sec-scattering}
We now investigate the scattering dynamics of giant atom and the bosonic field in the transmission line.
In the study of giant atoms, the long-time static scattering coefficients are often analyzed through the wave-function approach by calculating the eigenstates of total system \cite{Chen2022cp,Cai2021pra}.
%
%
In this section, we develop an alternative method to calculate the time-varying  scattering coefficients based on the linear response theory\cite{Nakajima1982book,Kubo1957JPSJ}. We first introduce a weak driving term in the Hamiltonian as a probe field, and then solve the time evolution of the responsed giant atom and the bosonic field. We divide the amplitude of the scattered bosonic fields by the probe field, and consequently derive the scattering coefficients.

\subsection{The probe drive}

As illustrated in Fig.~\ref{fig setup}(a),  an incident probe plane wave with frequency $\omega_d$ and amplitude $f$ is sent from the left to the right in the transmission line. The probe plane wave will then interact with all the coupling points of the giant atom in sequence. By setting the moment when the plane wave pass the leftmost coupling point as the initial time,  we write the total Hamiltonian as $H=H_s+H_D$ with the driving term $H_D$ given by
\begin{eqnarray}\label{HD}
H_D&=&
-\frac{i}{2}\hbar f \sum_m c_m e^{i\omega_d(t-\tau_m)}\Theta(t-\tau_m)\hat{b}+{\rm H.c.}=-\frac{i}{2}\hbar\Omega_f\hat{b}+{\rm H.c.}
\end{eqnarray}
Here, the parameter $\Omega_f \equiv f\sum_{m}c_me^{i\omega_d(t- \tau_m)}\Theta(t-\tau_m)$ describes the collective driving effect of all the coupling points, where $\tau_m\equiv
(x_m-x_1)/v$ is the traveling time of the drive and  $\Theta(x)$ is the Heaviside step function.

\subsection{Linear response of giant atom}

In order to solve the dynamics of giant atom in the presence of probe drive, we just need to replace the Hamiltonian $H_s$ in Heisenberg EOMs (\ref{EOMb-1}) and  (\ref{EOMb-2}) by $H=H_s+H_D$, and extend the solution of the giant atom operator $\hat{b}(t)$ given by Eq.~(\ref{Solution of EOM2}) to the following form
\begin{eqnarray}\label{bt}
 \hat{b}(t)&\equiv& \zeta(t)+\chi(t)\hat{b}(0)+\int_{-\infty}^\infty dk
\xi_k(t)\hat{a}_k(0).
\end{eqnarray}
Here, $\zeta(t)$ represents the linear response of the giant atom to the probe drive. Using Laplace transformation similar to Eqs.~(\ref{Solution of EOM3}) and (\ref{InverseEs}), we obtain the analytical solutions of $\zeta(t), \chi(t)$ and $\xi_k(t)$. The solutions of $\chi(t)$ and $\xi_k(t)$ are the same as Eqs.~(\ref{PolesI}) and (\ref{PolesII}) while the linear response term $\zeta(t)$ is given by
\begin{eqnarray}\label{zetat}
\zeta(t)&=&\frac{f^*\sum_mc^*_me^{-i\omega_d
(t-\tau_m)}}{2(-i\Delta+\Gamma\sum_{m,m'}c_mc_{m'}e^{i\omega_d|\tau_{mm'}|})}\nonumber\\
&&+\sum_{n}\frac{f^*\sum_mc^*_me^{s_n(t-
\tau_m)}}{2(s_n+i\omega_d)(1-\Gamma\sum_{m,m'}c_mc_{m'}|\tau_{mm'}|e^{-s_n|\tau_{mm'}|})},
\end{eqnarray}
where the parameter $\Delta\equiv\omega_d-\Omega$ is the atom-drive frequency detuning, and $s_n$ represents the roots of Eq.~(\ref{polesEq}).

\subsection{Scattering dynamics of bosonic field}
We now discuss the scattering dynamics of bosonic field in the transmission line. In order to calculate the scattering coefficients, we define the following field operator
\bea
\hat{\Psi}(x,t)
\equiv\int_{-\infty}^\infty dk\sqrt{\omega_k}
 e^{ikx}\ \hat{a}_k(t).
 \eea
Note that the above field operator $\hat{\Psi}(x,t)$ is different from the previous field operator $\hat{\varphi}(x,t)$ defined in Eq.~(\ref{numberdensity}) by a factor $\sqrt{\omega_k}$ in the integrand. As a result, the quantity of $|\langle\hat{\Psi}(x,t)\rangle|^2$ measures the intensity of field energy (scaled by $\hbar$) while the quantity of $|\langle\hat{\varphi}(x,t)\rangle|^2$ measures the density of bosonic mode number (or the probability for single photon/phonon dynamics) in the transmission line.
Given the solution of the giant atom operator $\hat{b}(t)$, we
have the time evolution of the field operator $\hat{\Psi}(x,t)$ from Eq.~(\ref{akt})
\begin{eqnarray}\label{Bigfieldoperator1}
\hat{\Psi}(x,t)
&=&-i\hbar \Gamma\sum_{m'}c_{m'}\hat{b}(t-|x-x_{m'}|/v)\Theta(t-|x-x_{m'}|/v)\nonumber\\
 &&+\int_{-\infty}^\infty
 dk\sqrt{\omega_k}
 e^{ikx_m}\ \hat{a}_k(0)e^{-i\omega_kt}.
\end{eqnarray}
Assuming the initial state of giant atom and the bosonic field are both in the ground states, we then have $\langle \hat{b}(t)\rangle=\zeta(t)$ according to Eq.~(\ref{bt}).
The reflection coefficient can be calculated from the
bosonic field energy intensity near the left side of the leftmost first coupling point $\langle\hat{\Psi}(x_1,t)\rangle$.
According to the driving Hamiltonian Eq.~(\ref{HD}),
the corresponding probe driving amplitude is $\frac{1}{2}i\hbar
f^*e^{-i\omega_dt}$.
Therefore, the dynamic of reflection coefficient is given by
\begin{eqnarray}\label{Rt}
\mathcal{R}(t)\equiv\frac{|\langle\Psi(x_1,t)\rangle|^2}{|\frac{1}{2}\hbar f^*e^{-i\omega_dt}|^2}
&=&\frac{4\Gamma^2}{|f^*|^2}|\sum_{m'}c_{m'}\zeta(t-\tau_{m'})|^2.
\end{eqnarray}
According to Eq.~(\ref{zetat}),
the corresponding long-time limit of  reflection coefficient is given by
\begin{eqnarray}\label{Rlim}
\mathcal{R}(t\to\infty)=\frac{\Gamma^2|\sum_{m}c_me^{i\omega_d\tau_m}|^4}{|-i\Delta+\Gamma\sum_{m,m'}c_mc_{m'}e^{i\omega_d|\tau_{mm'}|}|^2}.
\end{eqnarray}
In contrast, the field near the right side of the rightmost last coupling point is the superposition of the
probe field and transmitted field.
Assuming there are $N$ coupling points in total,
the dynamics of transmission
coefficient is then given by
\begin{eqnarray}\label{Tt}
\mathcal{T}(t)&\equiv&\frac{|\frac{1}{2}i\hbar f^*e^{-i\omega_d(t- \tau_N)}+\langle\hat{\Psi}(x_N,t)\rangle|^2}{|\frac{1}{2}\hbar f^*e^{-i\omega_dt}|^2}\nonumber\\
&=&\frac{|f^*e^{-i\omega_d(t- \tau_N)}-2\Gamma
\sum_{m'}c_{m'}\zeta(t-\tau_N+\tau_{m'})|^2}{|f^*e^{-i\omega_dt}|^2}
\end{eqnarray}
with corresponding long-time limit
\begin{eqnarray}\label{}
\mathcal{T}(t\to\infty)
=\Big|1-\frac{\Gamma|\sum_{m}c_me^{i\omega_d\tau_m}|^2}{-i\Delta+\Gamma\sum_{m,m'}c_mc_{m'}e^{i\omega_d|\tau_{mm'}|}}
\Big|^2.
\end{eqnarray}

\begin{figure}
\centering
\includegraphics[width=0.7\linewidth]{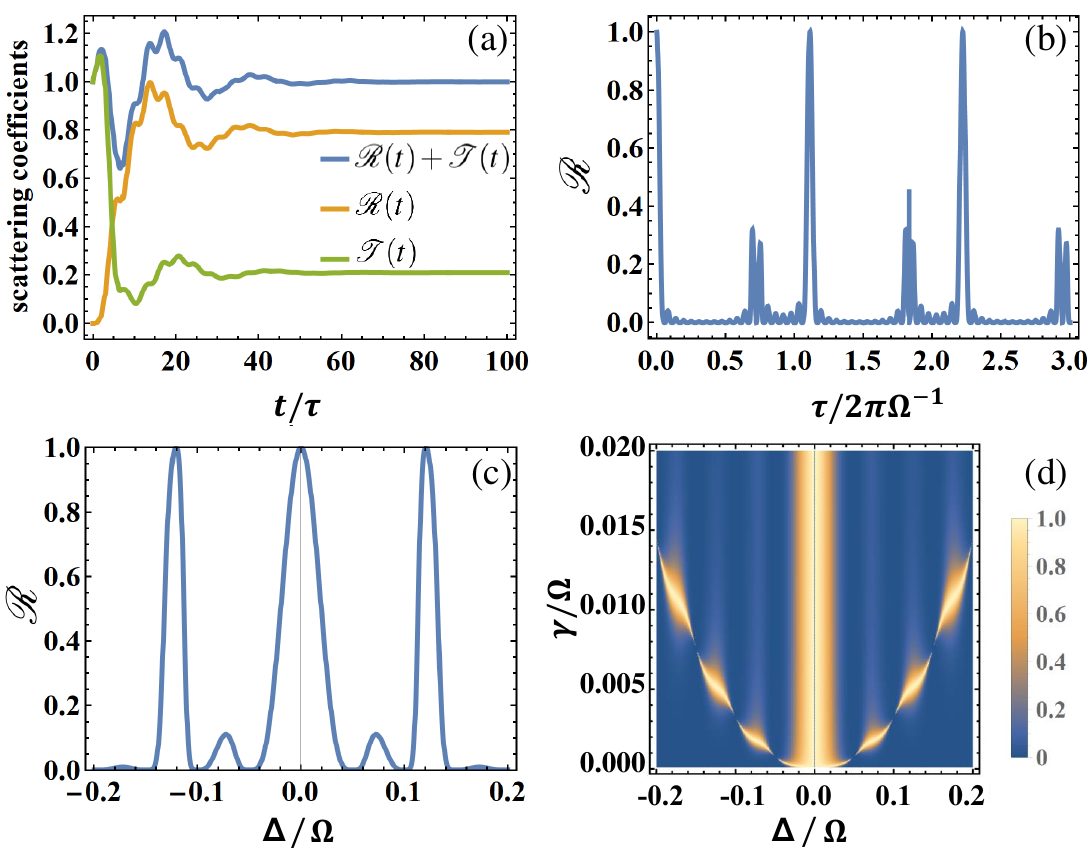}
\caption{{\bf Scattering coefficients for a giant atom with $N$ equal strength coupling point.}  {\bf (a)} The time-dependent reflection coefficient $\mathcal{R}(t)$ and transmission coefficient $\mathcal{T}(t)$. In the long time limit,   $\mathcal{R}(\infty)$ is consistent with the analytical result in Eq.(\ref{Reflect1}) satisfying  $\mathcal{R}(\infty)+\mathcal{T}(\infty)=1$. {\bf (b)-(c):} Stationary reflection coefficient $\mathcal{R}(\infty)$ as functions of  the delay time $\tau$ between the neighbouring coupling points (b);  the drive-atom frequency detuning  $\Delta=\omega_d-\Omega$ (c); and the relaxation rate $\gamma$ at single coupling point (d).
Parameters: (a) $N=5$, $\Omega\tau=2\pi$, $\omega_d=0.9\Omega$ and $\gamma=0.02\Omega$; (b) $N=20$, $\Omega=1, \omega_d=0.9\Omega$  and $\gamma=0.02\Omega$; (c) $N=20$, $\Omega\tau=2\pi$ and $\gamma=0.005\Omega$; (d) $N=20$ and $\Omega\tau=2\pi$.
\label{Fig-Reflection}}
\end{figure}

In Fig.~\ref{Fig-Reflection}(a), we plot the time-dependent reflection and transmission coefficients, calculated from Eqs.~(\ref{Rt}) and (\ref{Tt}), for a giant atom with $N=5$ coupling points. Note that, in the early time of scattering process, the sum of reflection and transmission can be either larger or smaller than one. This is because the giant atom has not yet established a static state due to the finite time delays among legs, and thus the coupling points of giant atom can store or release field during this transient process.
In the long-time limit when the giant atom establishes a static state, we have
\begin{eqnarray}\label{}
\mathcal{T}(\infty)
&=&1-\frac{2\Gamma^2|\sum_{m}c_me^{i\omega_d\tau_m}|^2}{|-i\Delta+\Gamma\sum_{m,m'}c_mc_{m'}e^{i\omega_d|\tau_{mm'}|}|^2}\mathrm{Re}[\sum_{m,m'}c_mc_{m'}e^{i\omega_d|\tau_{mm'}|}]\nn
&&+\frac{\Gamma^2|\sum_{m}c_me^{i\omega_d\tau_m}|^4}{|-i\Delta+\Gamma\sum_{m,m'}c_mc_{m'}e^{i\omega_d|\tau_{mm'}|}|^2}\nonumber\\
&=&1-\frac{\Gamma^2|\sum_{m}c_me^{i\omega_d\tau_m}|^4}{|-i\Delta+\Gamma\sum_{m,m'}e^{i\omega_d|\tau_{mm'}|}|^2}=1-\mathcal{R}(\infty).
\end{eqnarray}
 Here, we have used the following relationship
 \begin{eqnarray}\label{}
|\sum_{m}c_me^{i\omega_d\tau_m}|^2&&=\sum_{m,m'}c_mc_{m'}e^{i\omega_d(\tau_{m}-\tau_{m'})}={\rm Re}[\sum_{m,m'}c_mc_{m'}e^{i\omega_d|\tau_{mm'}|}].
\end{eqnarray}

\subsection{Examples}

As a concrete example, we consider a $N$-leg giant atom with equal coupling strength $c_m=c_0$ and equal delay time $\tau$ between the neighboring coupling points.
Therefore, all the possible delay times are
$\tau_{mm'}=n\tau$ with $n=0,1,2,\cdot\cdot\cdot,N-1$.
The combination number
of each possible delay time is $N$ for $n=0$; and $2(N-n)$ for
$n\neq 0$.
From Eq.~(\ref{Rt}) and the identity
\begin{eqnarray}\label{identity1}
\sum_{n=0}^{N-1}(N-n)e^{in\phi}=\frac{N}{2}+\frac{1-\cos
(N\phi)}{2(1-\cos \phi)}+i\frac{N\sin\phi-\sin (N\phi)}{2(1-\cos
\phi)},
\end{eqnarray}
we have the analytical expression of the stationary reflection coefficient for the $N$-leg model
\begin{eqnarray}\label{Reflect1}
\mathcal{R}(\infty)
&=&\frac{\Big[\frac{\gamma}{2}\frac{1-\cos(N\omega_d\tau)}{1-\cos(\omega_d\tau)}\Big]^2}
{\Big[\omega_d-\Omega-\frac{\gamma}{2}\frac{N\sin(\omega_d\tau)-\sin(N\omega_d\tau)}{1-\cos(\omega_d\tau)}\Big]^2+\Big[\frac{\gamma}{2}\frac{1-\cos(N\omega_d\tau)}{1-\cos(\omega_d\tau)}\Big]^2
}.
\end{eqnarray}
Here, $\gamma=2\Gamma c_0^2$ is the relaxation rate the single coupling point.
For $N$-leg giant atom with alternative couplings sequence $c_m=(-1)^{m}c_0$, the reflection coefficient  can be derived  by replacing the $\omega_d \tau$ in Eq.(\ref{Reflect1}) with $\omega_d \tau+\pi$.
From the above expression of reflection,
 the bare atomic level is modified by a Lamb shift $\Delta_L$ and an effective decay rate $\tilde{\gamma}$ given by
\bea\label{eq-DeltL}
\Delta_L=\frac{\gamma}{2}\frac{N\sin(\omega_d\tau)-\sin(N\omega_d\tau)}{1-\cos(\omega_d\tau)},\ \ \ \tilde{\gamma}=\frac{\gamma}{2}\frac{1-\cos(N\omega_d\tau)}{1-\cos(\omega_d\tau)}.
\eea
When the driving field is
resonant with this shifted level, i.e.,
\bea\label{eq-sidepeaks}
\omega_d=\Omega+\frac{\gamma}{2}\frac{N\sin(\omega_d\tau)-\sin(N\omega_d\tau)}{1-\cos(\omega_d\tau)},
\eea
the reflection coefficient is $\mathcal{R}=1$. The full reflection condition can be obtained by adjusting the frequency of the giant atom $\Omega$, the driving frequency detune $\Delta$, and the relaxation rate $\gamma$ at single coupling point.
%
%

In Figs.~\ref{Fig-Reflection}(b)-(d), we plot the long-time limit of reflection coefficient of the giant atom with $N=20$ coupling points.
Note that the Lamb shift and the effective decay rate given by Eq.~(\ref{eq-DeltL}) are different from the results obtained in the absence of driving field, cf., Eqs.~(\ref{EOM2IDT-3a}) and (\ref{EOM2IDT-3b}), by replacing the atomic frequency $\Omega$ with the driving frequency $\omega_d$.
In fact, the propagating phase accumulation can be divided into $\omega_d\tau=\Omega\tau+\Delta\tau$, where the second phase accumulation from detuning is neglected in the Markovian regime \cite{chen2023giantatom}. Such a detuning-dependent phase is origin of peculiar  scattering features in the transmission and reflection spectra \cite{du2021prr}.

\subsection{Non-Markovian giant atom}
An interesting phenomenon is that the giant atom can have multiple full reflection points as the driving frequency tunes.
As shown in Fig.~\ref{Fig-Reflection}(d), for point-like small atom ($\gamma\tau=0$), the full reflection happens only for $\omega_d=\Omega$. However, as $\gamma$ increases, two more side full-reflection peaks appear for non-resonant driving frequencies. To illustrate how to search for the full-reflection side peaks, we consider the resonant condition Eq.~(\ref{eq-sidepeaks}) for the case of $\Omega\tau=2\pi$,
\bea
\Delta=\frac{\gamma}{2}\frac{N\sin(\Delta\tau)-\sin(N\Delta\tau)}{1-\cos(\Delta\tau)}.
\eea
\begin{figure}
\centering
\includegraphics[width=0.6\linewidth]{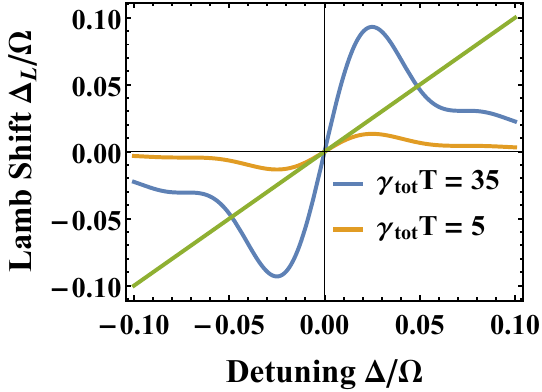}
\caption{Lamb shift $\Delta_L$ as a function of frequency detuning $\Delta=\omega_d-\Omega$ for the Markovian (orange) and non-Markovian (blue) giant atoms. The green line indicates the resonance condition $\Delta_L=\Delta$ for point-like small atom.
Parameters:  $N=20$, $\Omega\tau=2\pi$, $T=(N-1)\tau$ and $\gamma_{tot}=N^2\gamma$. \label{Fig-LS}}
\end{figure}
In Fig.\ref{Fig-LS}, we plot the two sides of the above condition as functions of frequency detuning $\Delta=\omega_d-\Omega$. Except the trivial solution $\Delta=0$, the existence of other solutions is
\begin{eqnarray}\label{gammac}
\frac{d}{d\Delta}\Big|_{\Delta=0}\Big(\frac{\gamma}{2}\frac{N\sin(\Delta\tau)-\sin(N\Delta\tau)}{1-\cos(\Delta\tau)}\Big)=\frac{\gamma}{6}N(N^2-1)\tau>1.
\end{eqnarray}
By introducing the total decay rate $\gamma_{tot}\equiv
N^2\gamma$ and the total delay time $T\equiv(N-1)\tau$, the above condition becomes
\begin{eqnarray}\label{eq-nonM}
\gamma_{tot} T>\frac{6}{1+1/N}.
\end{eqnarray}
We see that the full-reflection side peaks only appear when the total delay time of field across the all the coupling points is NOT negligible.
In fact, the condition Eq.~(\ref{eq-nonM}) can be taken as a criteria to define the non-Markovian regime for a giant atom (in the case of $\Omega\tau=2\pi$).

\section{Catch and release of propagating field: a photon/phonon tweezer}\label{sec-tweezer}

 In quantum networks, the control of traveling quantum fields encoded with information (flying qubit)  is crucial for coherent information transfer \cite{li2022prb,li2022automatica}. The quantum information between distant nodes is physically shared by catching and releasing flying qubits. In this section, we propose a \textit{photon/phonon tweezer} with two-leg giant atom that can catch and release propagating bosonic field in the transmission line.

\subsection{Dark state and bound state }
%
%

Suppose that the giant atom is in the excited state and the environment is in the vacuum state initially.
According to Eqs.~(\ref{Solution of EOM2}) and (\ref{PolesI}), the atom does not decay totally to zero if there exists purely imaginary pole $s_n$, and will stay in the \textit{dark state} decoupled completely to the dissipative environment.
%
%
%
Assuming the pure imaginary pole of Eq.~(\ref{polesEq}) has the form  $s_n=-i\frac{2n\pi}{N\tau},\  n\in{\rm Z}$,  we have the dark state condition for the special case that all the coupling points have equal coupling strength $c_m=c_0\  (m=1,2,\cdots, N)$
 \cite{guo2020prr}
\begin{eqnarray}\label{resonantdark1}
\Omega\tau=\frac{2n\pi}{N}-\frac{1}{2}N\gamma\tau\cot\Big(\frac{n\pi}{N}\Big), \ \ \ \ n\in{\rm Z}.
\end{eqnarray}
Under this condition, the dynamics of atomic dark state is
\begin{eqnarray}\label{evolutiondark1}
\beta(t)
=A(n)e^{-i\frac{2n\pi}{N\tau}t}\ \ \ \mathrm{with}\ \ \  A(n)\equiv\frac{2\sin^2(n\pi/N)}{2\sin^2(n\pi/N)+N\gamma\tau}.
\end{eqnarray}
In the case that the coupling sequence is set to be $c_m=(-1)^mc_0 \  (m=1,2,\cdots, N)$, we obtain the  dark state condition
\begin{eqnarray}\label{resonantdark2}
\Omega\tau=\frac{2n\pi}{N}-\frac{1}{2}N\gamma\tau\tan\Big(\frac{n\pi}{N}\Big), \ \ \ \ n\in{\rm Z}.
\end{eqnarray}
The corresponding dynamics of atomic dark state  is given by
\begin{eqnarray}\label{evolutiondark}
\beta(t)
=\frac{2\cos^2(n\pi/N)}{2\cos^2(n\pi/N)+N\gamma\tau}e^{-i\frac{2n\pi}{N\tau}t}.
\end{eqnarray}
According to Eqs.~(\ref{evolutiondark1}) and (\ref{evolutiondark}), the probability amplitude of giant atom is less than one $|\beta(t)|<1$ in the non-Markovian regime ($\gamma\tau>0$). This is because giant atom needs a finite time to build the destructive interference among different coupling points. During this transient time, some field excitation  leaks outside the coupling points  and never comes back.

The difference between the Markovian and  the non-Markovian giant atom can also be revealed  by the bound state trapped inside the coupling points. From Eq.~(\ref{numberdensity}), we have the field function in the transmission line as follows
\begin{eqnarray}\label{fieldnumA}
\varphi(x,t)
&\equiv&\langle\hat{\varphi}(x,t)\rangle\approx\frac{1}{\sqrt{2\pi}}\int_{-\infty}^\infty
 dk
 e^{i(kx-\omega_kt)}\langle\hat{a}_k(0)\rangle\nonumber\\
 && -i\sqrt{\frac{\Gamma}{v}}\sum_{m'}c_{m'}\beta(t-|x-x_{m'}|/v)\Theta(t-|x-x_{m'}|/v).
\end{eqnarray}
In the case of equal coupling strength $c_m=c_0\  (m=1,2,\cdots, N)$ and for a given dark mode $s_n = - i\frac{2 n \pi}{N \tau}$, the corresponding stationary field number density $p_n(x)=|\varphi(x,t\rightarrow+\infty)|^2$ can be directly calculated from Eqs.~(\ref{fieldnumA}) and (\ref{evolutiondark1}) (see detailed derivation in Ref.~\cite{guo2020prr})
\bea\label{eq-pnxN}
p_n(x)= \frac{8 \gamma}{v} \frac{\sin^2 \frac{n \pi}{N}\sin^2 \Big( \frac{n \pi}{N} m' \Big)}{\Big( 2 \sin^2 \frac{n \pi}{N} + N \gamma \tau \Big)^2}
 \sin^2 \Big[ \frac{n \pi}{N} \Big( m' + 2 \lambda - 1 \Big) \Big].
\eea
Here, we have parametrized the position coordinate as $x = (m' - 1) v \tau + \lambda v \tau$ with $m' = 1, 2, \ldots, N$  and $\lambda \in [0,1)$. Also, note that $\Gamma=\frac{2\pi\Omega}{\hbar^2 v}$ and $\gamma=2\Gamma c^2_0$. When the position $x$ is outside the interval $[x_1, x_m]$, the field intensity completely vanishes. Clearly, the field intensity is zero at the two ends, $x_1 = 0$ (i.e., $m' = 1$ and $\lambda = 0$) and $x_N = (N - 1) v \tau$ (i.e., $m' = N$ and $\lambda = 0$). This is in stark contrast to the bound state due to local impurity (small atom) in the environment, where the field  bound state is exponentially suppressed on both sides of impurity (small atom). The total field excitation of bound state can be calculated analytically \cite{guo2020prr}
\bea
\label{PT}
I(n) &=& \int_{x_1}^{x_N} p_n (x) \ dx
= \frac{2 N \gamma \tau \sin^2 \Big( \frac{n \pi}{N} \Big)}{\Big[ 2 \sin^2 \big( \frac{n \pi}{N}\big) + N \gamma \tau  \Big]^2} \Big[ 1 + \frac{N}{4 n \pi} \sin \Big( \frac{2 n \pi}{N} \big) \Big].
\eea
Note that the total field excitation disappears $I(n)\rightarrow 0$ in the Markovian limit $\gamma\tau\rightarrow 0$. As a result, the bound state can only exist in the non-Markovian regime $\gamma\tau > 0$.

Compared to the dark state condition in Markovian limit $\gamma\tau\rightarrow 0$, the dark state condition for a non-Markovian giant atom given by Eq.~(\ref{resonantdark1}) or (\ref{resonantdark2})
is nonlinear with respect to the dark mode integer $n$. Therefore, there is possibility of existing two dark modes $s_{n_1}=\frac{2\pi n_1}{N\tau}$ and $s_{n_2}=\frac{2\pi n_2}{N\tau}$ with $n_1\neq n_2$ for the same system parameter setting. For example, we can choose  the following system parameters to satisfy the dark state condition Eq.~(\ref{resonantdark1}) \cite{guo2020prr}
\bea
\label{bidarks3}
\mleft\{ \begin{array}{l}
\Omega \tau = \frac{2 n_1 \pi}{N} - \frac{2 (n_1 - n_2) \pi}{N} \frac{\cot\mleft( \frac{n_1 \pi}{N} \mright)}{\cot \mleft( \frac{n_1 \pi}{N} \mright) - \cot \mleft( \frac{n_2 \pi}{N} \mright)} > 0,
\\
\gamma \tau = \frac{4 (n_1 - n_2) \pi}{N^2} \frac{1}{\cot \mleft( \frac{n_1 \pi}{N} \mright) - \cot \mleft( \frac{n_2 \pi}{N} \mright)} > 0.
\end{array} \mright.
\eea
In this case, the long-time dynamics of the giant atom is $\beta (t) \rightarrow A (n_1) e^{- i \Omega_{n_1} t} + A (n_2) e^{- i \Omega_{n_2} t}$ with $A(n)$ given by Eq.~(\ref{evolutiondark1}). 
The excitation of atom $|\beta(t)|^2$ and the total field of two bound states $I(n_1,n_2)\equiv\int p(x,t\rightarrow \infty)dx$ in the environment are given by \cite{guo2020prr}
\bea
\label{}
\mleft\{ \begin{array}{l}
\abssq{\beta (t)} = A^2 (n_1) + A^2 (n_2) + 2 A (n_1) A (n_2) \cos \mleft[ \mleft( \Omega_{n_1} - \Omega_{n_2} \mright) t \mright],
\\
I (n_1, n_2) = I (n_1) + I (n_2) - 2 A (n_1) A (n_2) \cos \mleft[ \mleft( \Omega_{n_1} - \Omega_{n_2} \mright) t \mright].
\end{array} \mright.
\eea
Because the oscillating bound state does not decay, total excitation probability of the atom and the field $\abssq{\beta (t)} +I(n_1,n_2)$ is conserved,

\begin{figure}
\centering
\includegraphics[width=0.7\linewidth]{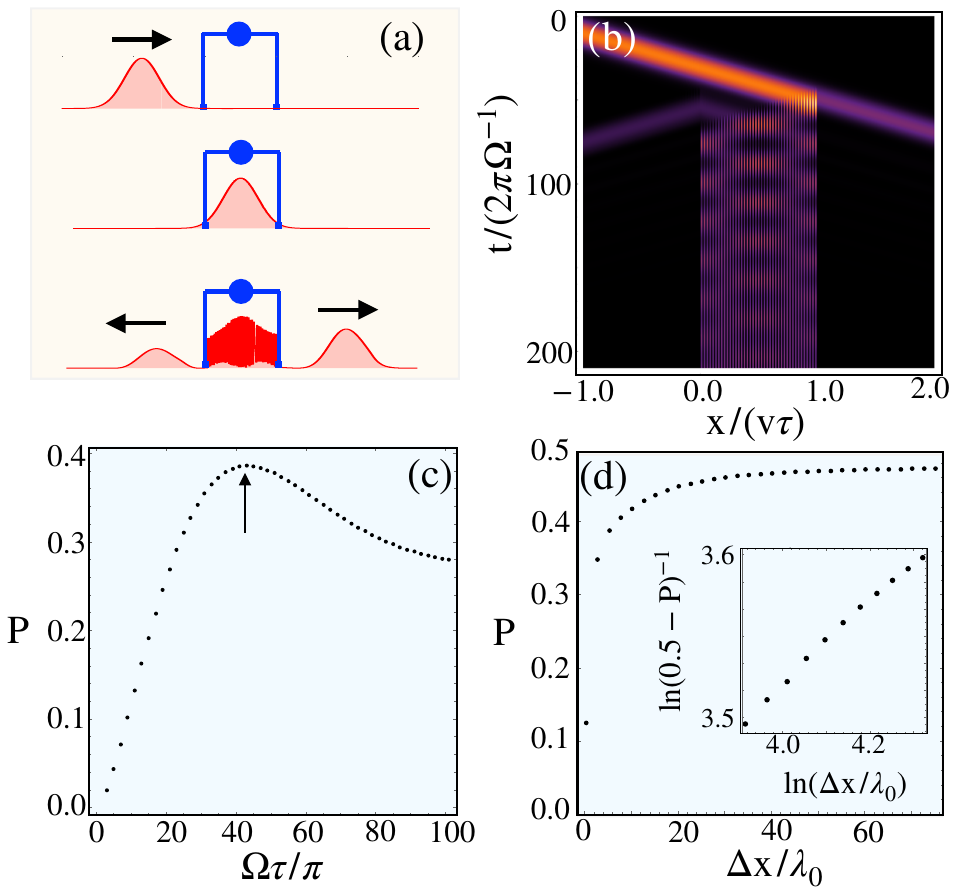}
\caption{{\bf (a)} Three-steps protocol to catch a wave packet propagating in the transmission line: (upper) the atom's frequency $\Omega$ is tuned far away from the centre frequency of wave packet $\Omega\gg \Omega_0$; (middle) the atom's frequency is turned on the resonance condition $\Omega=\Omega_0$ and dark condition $\Omega\tau/\pi=2n+1$; (lower) wave packet is caught with probability $P$ meanwhile also reflected and transmitted. {\bf (b)} Time evolution of field intensity $|\varphi(x,t)|^2$, cf. Eq.~(\ref{eq-fieldfun}),  in the transmission line corresponding to figure (a). {\bf (c)} The catching probability $P$ as a function of the phase difference between giant atom's two legs $\Omega\tau/\pi=2n+1$ with arrow indicating the optimal catching condition of $n\approx 4\Delta x/\lambda_0$. The width of wave packet is set to be $\Delta x/\lambda_0=5$ in the simulation. {\bf (d)} The catching probability $P$ as a function of wave packet width $\Delta x/\lambda_0$ on the optimal point $\Omega\tau/\pi=8\Delta x/\lambda_0+1$. The inset is the log-log plot of $\ln(1/2-P)$ versus $\ln(\Delta x/\lambda_0)$ showing that the asymptotic behaviour of $P$ follows a power law $P=0.5-C(\Delta x/\lambda_0)^{-\alpha}$ with fitting exponent $\alpha\approx 0.2$.
\label{Fig-PT1}}
\end{figure}

\subsection{Catching a single propagating wave packet}

As discussed in the above section, the bound state of bosonic field in the transmission line only exists for non-Markovian giant atom. In fact, this non-Markovianity can be explored as a resource to realize a tweezer that can catch and release a propagating field in the transmission line that is useful for quantum signal processing \cite{hann2019prl}.

As shown in Fig.~\ref{Fig-PT1}(a), we assume a single bosonic wave packet propagating from left to right in the transmission line with the free field function described by
\begin{eqnarray}\label{varphi0xt}
\varphi_0(x,t)&=&\frac{1}{\sqrt{2\pi}}\int_{-\infty}^{\infty}dke^{ik(x-vt)}\phi_k(0)\nonumber\\
&=&\Bigg(\frac{2\sigma^2_k}{\pi}\Bigg)^{\frac 1 4}\exp\Big[-\sigma^2_k(x-x_0-vt)^2+ik_0(x-x_0-vt)\Big].
\end{eqnarray}
Here, the coefficient $\phi_k(0)=\langle \hat{a}_k(0)\rangle$ follows a Gaussian distribution around a central wavenumber $k_0>0$ with standard deviation $\sigma_k$ in the momentum $k$ space,
\begin{eqnarray}
\phi_k(0)=\Bigg(\frac{1}{2\pi\sigma^2_k}\Bigg)^{\frac 1 4}\exp\Big[{-\frac 1 4\Big(\frac{k-k_0}{\sigma_k}\Big)^{2}-ikx_0}\Big].
\end{eqnarray}
The field intensity $|\varphi_0(x,t)|^2$ indeed represents a propagating wave packet with central position $x=x_0+vt$ and width $\Delta x=(2\sigma_k)^{-1}$. According to Eq.~(\ref{Solution of EOM2}) including the initial field condition, we have the atomic EOM of $\beta(t)=\langle\hat{b}(t)\rangle$ as follows
\begin{eqnarray}\label{}
\frac{d}{dt}\beta(t)
&=&-i\Omega \beta(t)-\frac 1 2 N\gamma b(t)-\gamma\sum^{N-1}_{l=1}\beta(t-l\tau) \Theta(t-l\tau)\nn
&&-i\sqrt{\frac{\gamma v}{2} }\sum_{m = 1}^N\varphi_0(x_m,t),
\end{eqnarray}
and the field EOM in the transmission line from Eq.~(\ref{fieldnumA}),
\begin{eqnarray}\label{eq-fieldfun}
\varphi(x,t)&=&\varphi_0(x,t)-i\sqrt{\frac{\gamma }{2v}}\sum_{m = 1}^N\beta\Big(t-\frac{|x-x_m|}{v}\Big) \Theta\Big(t-\frac{|x-x_m|}{v}\Big).
\end{eqnarray}

In Fig.~\ref{Fig-PT1}(a), we  propose a three-step protocol to catch the propagating wave packet with a two-leg giant atom. The initial field in the transmission line is given by Eq.~(\ref{varphi0xt}), and the giant atom is in the ground state. The upper subfigure shows a wave packet in the transmission propagates from left side of giant atom with centre frequency $\Omega_0=2\pi/k_0$ and wave packet width $\Delta x=5\lambda_0$, where $\lambda_0=2\pi/k_0$ is the centre wavelength of wave packet.
At the beginning, the characteristic frequency $\Omega$ of giant atom is tuned far away from the wave packet central frequency, i.e., $\Omega\gg\Omega_0$.
When the wave packet arrives at the center of atom's two legs as shown by the middle subfigure, the atom's frequency is turned on resonance with the wave packet's frequency ($\Omega=\Omega_0$) and also satisfies the dark state condition, i.e., $\Omega\tau=(2n+1)\pi$ with $n\in{\rm Z}$.
After that, as shown by the lower subfigure, the wave packet can be caught or trapped by the giant atom with probability $P$, meanwhile the wave packet is also reflected and transmitted.

In Fig.~\ref{Fig-PT1}(b), we show the time evolution of field intensity $|\varphi_0(x,t)|^2$ in the transmission line corresponding to Fig.~\ref{Fig-PT1}(a). In Fig.~\ref{Fig-PT1}(c), we plot the catching probability $P$ as a function of the phase difference $\Omega\tau/\pi=2n+1$ between giant the atom's two legs. We see that the optimal choice of phase difference to catch the wave packet is given approximately by $n\approx 20$, which is about four times multiple of the wave packet width measured by the centre wavelength,
 i.e., $n= 4\Delta x/\lambda_0$ or equivalently
\bea\label{eq-opt}
\frac{\Omega\tau}{\pi}=\frac{8\Delta x}{\lambda_0}+1.
\eea
This above empirical optimal condition is verified by numerical simulations for other wave packet lengths. In Fig.~\ref{Fig-PT1}(d), we plot the  catching probability $P$ as a function of wave packet width $\Delta x/\lambda_0$ on the optimal condition Eq.~(\ref{eq-opt}).
We see that the optimal catching probability approaches $P=50\%$ as the wave packet width increases. Assuming the asymptotic behaviour follows a power-law, i.e., $P=0.5-C(\Delta x/\lambda_0)^{-\alpha}$, the inset shows the power-law exponent is $\alpha\approx 0.2$. In principle, the optimal catching probability  can reach $P=50\%$ as the wave packet approaches a plane wave (the distance of giant atom's two legs becomes infinity).

\begin{figure}
\centering
\includegraphics[width=1.0\linewidth]{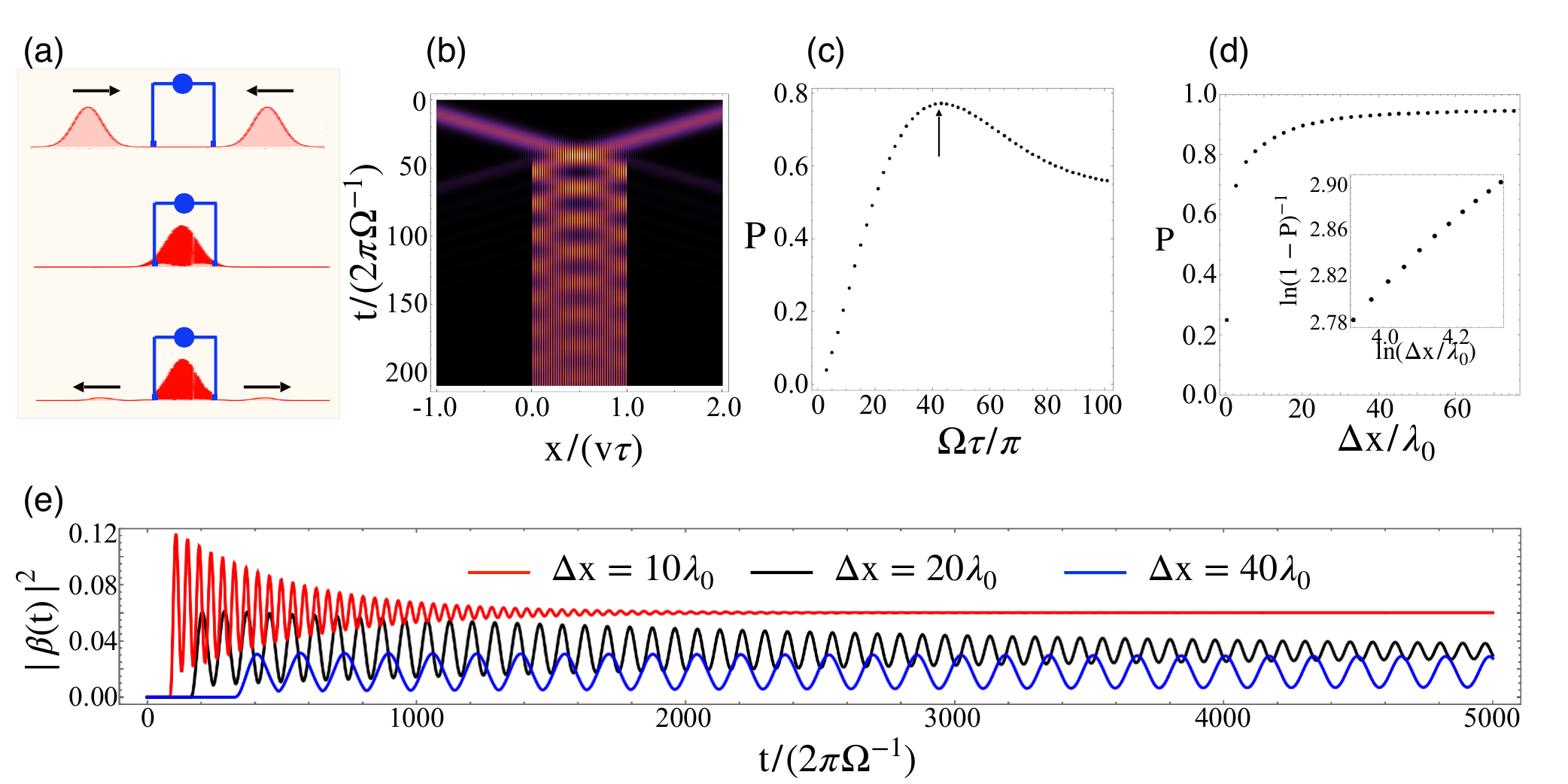}
\caption{{\bf (a)} Three steps to catch a superposition state of two counter-propagating wave packets in the transmission line: (upper) free propagating of two wave packets by setting the frequency of giant atom $\Omega\gg \Omega_0$; (middle) the frequency of giant atom is turned to satisfy the resonance condition $\Omega=\Omega_0$ and darks state condition $\Omega\tau/\pi=2n+1$; (lower) two wave packets are caught with probability $P$ while some are lost indicated by arrows. {\bf (b)} Time evolution of field intensity $|\varphi(x,t)|^2$, cf. Eq.~(\ref{eq-fieldfun}),  in the transmission line corresponding to figure (a). {\bf (c)} The catching probability $P$ as a function of the phase difference  between two legs $\Omega\tau/\pi=2n+1$. The arrow indicates the optimal catching condition: $n\approx 4\Delta x/\lambda_0$. The widths of wave packets are set to be $\Delta x/\lambda_0=5$ in the simulation. {\bf (d)} The catching probability $P$ as a function of wave packet width $\Delta x/\lambda_0$ on the optimal catching condition: $\Omega\tau/\pi=8\Delta x/\lambda_0+1$. The inset is the log-log plot of $\ln(1-P)$ versus $\ln(\Delta x/\lambda_0)$ showing that the asymptotic behaviour of $P$ follows a power law $P=1-C(\Delta x/\lambda_0)^{-\alpha}$ with fitting exponent $\alpha\approx 0.2$. {\bf (e)} The atomic excitation $|\beta(t)|^2$ for different wave packet widths as a function of time during the whole catching process.
\label{Fig-PT2}}
\end{figure}

\subsection{Catching superposition state of two counter-propagating wave packets}

In order to increase the probability to catch the propagating bosonic field, we propose to use another counter-propagating wave packet as shown in Fig.~\ref{Fig-PT2}(a). The initial field function is set to be in the superposition state of two counter-propagating wave packets as follows
\begin{eqnarray}\label{doublewave}
\varphi_0(x,t)&=&\Big(\frac{1}{8\pi\Delta x^2}\Big)^{\frac 1 4}e^{-\frac{1}{4\Delta x^2}(x-x_0-vt)^2+ik_0(x-x_0-vt)}\nonumber\\
&&+\Big(\frac{1}{8\pi\Delta x^2}\Big)^{\frac 1 4}e^{-\frac{1}{4\Delta x^2}(x+x_0-v\tau+vt)^2+ik_0(x+x_0-v\tau+vt)}.
\end{eqnarray}
The catching process is similar to that for the single wave packet, namely, the resonance and dark state condition is turned on when the two wave packets meet in the center of two legs.

In Fig.~\ref{Fig-PT1}(b), we show the time evolution of field intensity $|\varphi(x,t)|^2$, cf. Eq.~(\ref{eq-fieldfun}),  in the transmission line corresponding to Fig.~\ref{Fig-PT1}(a). In Fig.~\ref{Fig-PT2}(c), we plot the catching probability $P$ as a function of the phase difference $\Omega\tau/\pi=2n+1$ between giant atom's two legs. The arrow in this figure indicates the same optimal catching condition Eq.~(\ref{eq-opt}) for the single wave packet case, i.e., $\Omega\tau/\pi=8\Delta x/\lambda_0+1$. Fig.~\ref{Fig-PT2}(d) shows the optimal catching probability $P$ as a function of wave packet width $\Delta x/\lambda_0$ showing that the optimal probability approaches $P=100\%$ as the widths of two wave packets increase. The inset shows a power-law behavior $P=1-C(\Delta x/\lambda_0)^{-\alpha}$ with the power-law exponent $\alpha\approx 0.2$. The catching probability is already $P\approx95\%$ for the wave packet width $\Delta x=75\lambda_0$. In principle, the optimal catching probability can be $P=100\%$ as the two wave packets approach plane waves.
Note that, for a two-level giant atom, the the initial field state given by Eq.~(\ref{doublewave}) represents a single bosonic field excitation in the transmission line. If we want to catch two or more bosonic excitations, the two-level atom can be in replaced by a harmonic oscillator (LC circuit) with characteristic frequency $\Omega$ \cite{guo2020prr}.

In Fig.~\ref{Fig-PT2}(e), we plot the time evolution of atomic excitation for different wave packet widths during the whole catching process. We see that the atomic excitation is indeed very low, i.e., $|\beta(t)|^2 \leq 0.1$ for $\Delta x>10\lambda_0$, due to the destructive interference of two wave packets trapped between two legs.
Another interesting feature shown by Fig.~\ref{Fig-PT2}(e) is that the atomic excited probability $|\beta(t)=\langle\hat{b}(t)\rangle|^2$ oscillates longer as the width of wave packets increases. This can be explained by Eq.~(\ref{Solution of EOM2}) together with solutions Eqs.~(\ref{PolesI}) and (\ref{PolesII}). The solution for $\beta(t)$ is basically the superposition of the dark mode $s_n$, i.e., the first term on the right hand side of Eq.~(\ref{PolesII}), plus the residual harmonic modes from the propagating wave packet, i.e., the second term on the right hand side of Eq.~(\ref{PolesII}).
As result, the time evolution of $|\beta(t)|^2$ exhibits oscillations as time. However, the second term on the right hand side of Eq.~(\ref{PolesII}) in general dephases since it is a sum of many independent modes. Therefore, in the infinite long time limit, the contribution of $\beta(t)$ will be dominant by the single dark mode $s_n$ resulting a constant $|\beta(t\rightarrow\infty)|^2$. The dephasing time becomes longer if the second term on the right hand side of Eq.~(\ref{PolesII}) contains less modes (i.e., wider the width of wave packets), which is exactly shown by Fig.~\ref{Fig-PT2}(e). According to field function given by Eq.~(\ref{eq-fieldfun}), the bound state in the waveguide tweezed by the two-leg giant atom is also oscillating with some dephasing time, which can be seen from the plots of Fig.~\ref{Fig-PT1}(b) and Fig.~\ref{Fig-PT2}(b).

\begin{figure}
\centering
\includegraphics[width=\linewidth]{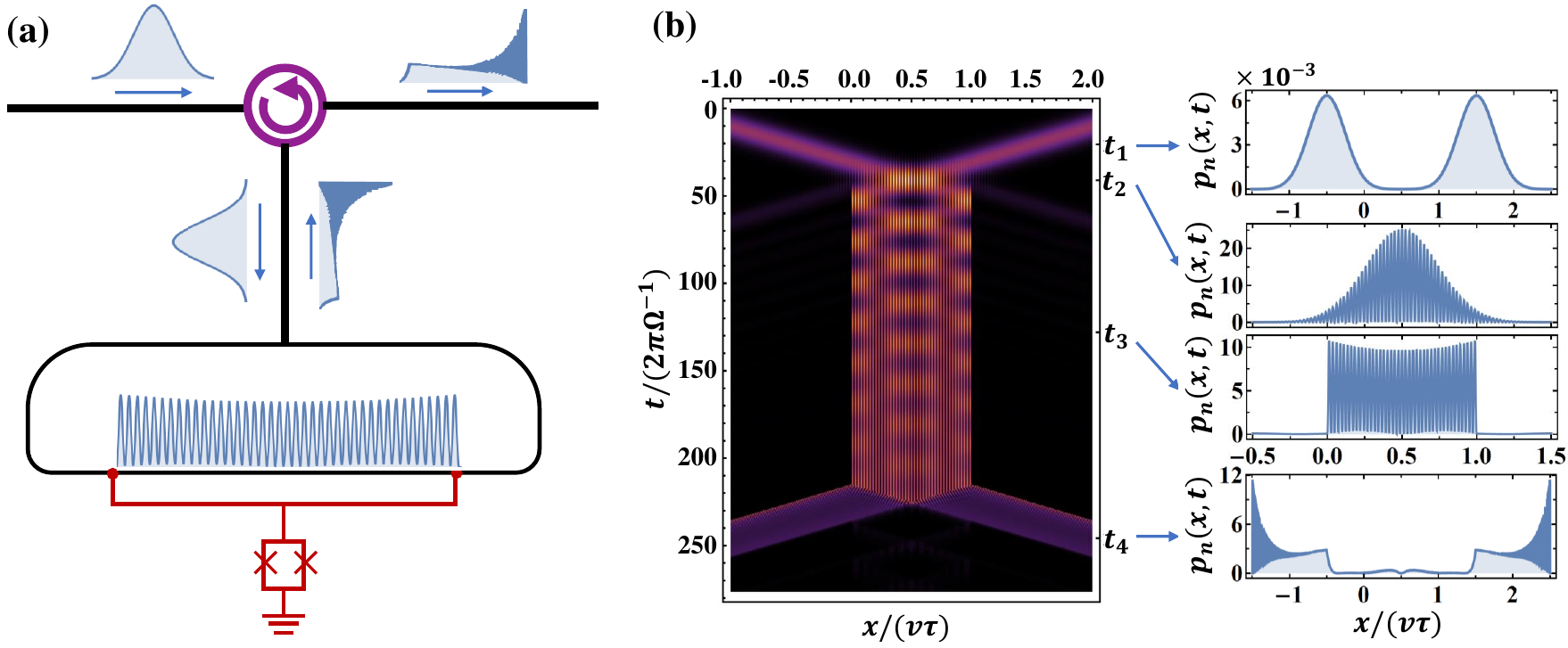}
\caption{ {\bf (a)} Circuit-QED implementation of catching and releasing a microwave photon with two-leg giant atom. {\bf (b)} Spacetime plot of the field intensity $|\varphi(x,t)|^2$, cf. Eq.~(\ref{eq-fieldfun}), in the loop transmission line during the whole catching and releasing process. The four subfigures on the right are the snapshots of field intensity at selected time moments, at which the two wave packets are separated  ($ t_1=41\pi/\Omega$), overlapped ($ t_2=82\pi/\Omega$), trapped ($t_3=256\pi/\Omega$), and  released ($t_4=491.5\pi/\Omega$). .
\label{fig-loop}}
\end{figure}

\subsection{Circuit-QED implementation}

As discussed above, the probability to catch a superposition state of two counter-propagating wave packets is much higher than the probability to catch a single propagating wave packet. Based on this fact, we propose a circuit quantum electrodynamics (QED) architecture, as shown in Fig.~\ref{fig-loop}(a), to implement our proposal.

In our scheme, the microwave photon in the form of single propagating wave packet is sent from the left side to the transmission line. The traveling photon is then routed downside via a circulator to another transmission line with a loop in the end. At the bifurcation of the loop, the single propagating wave packet is split into a superposition state of two counter propagating wave packets in the loop. Note that, to guarantee full transmission of field at the bifurcation point, the impedance of the loop line has to be twice of the impedance of incoming transmission line, which can be achieved by setting the width of the loop to be half of the incoming transmission line. The giant artificial atom made of tunable transmon is coupled to the loop line with two legs. When the two counter propagating wave packets meet in the middle of the loop, the giant atoms is turned to satisfy the dark state condition and thus the microwave photon is trapped between the two legs of giant atom. After the microwave photon is caught, as indicated by Fig.~\ref{Fig-PT2}(b), the shape of the tweezed photon will continue evolving to a stationary standing wave with $n$ antinodes, cf., Eq.~(\ref{eq-pnxN}) for $N=2$. When the photon is released again on demand, it will split into two separated propagating waves that have quite different shape from the incoming wave packet (transformation of flying qubits). They meet again at the bifurcation of the loop and emerge into a single traveling photon, which is then routed to the right transmission line.
In Fig.~\ref{fig-loop}(b), we show time evolution of  the field in the transmission line during the whole catching and releasing process. The four subfigures on the right are the snapshots of the field intensity distribution at selected time moments.

In the conventional circuit QED, the photon state can be caught and stored in a cavity or resonator with tunable coupling strength to the transmission line outside \cite{yin2013prl,pierre2014apl,dai2023pra}. In contrast,  our proposed scheme directly catches and stores the traveling field on the transmission line with two-leg giant atom. Different from the previous designs, we utilize non-Markovianity as a valuable resource allowing simpler and flexible circuit QED architecture, e.g., we can tune the bound state condition (\ref{resonantdark1}) to catch and release traveling photon with different frequencies.

Note that  the shape of released field is different from that of incident field pulse. In our current proposal,  the qubit information is encoded with the vacuum and single-excitation states as others \cite{li2022prb,li2022automatica} while the shape of the pulse is not used to encode information.

\section{Summary and Outlook}\label{sec-sumlook}

In this work, we have investigated the dynamics of propagating bosonic field coupled to a giant atom. We started from the model of a giant atom that interacts with the bosonic modes in the transmission line via multiple coupling legs stretching from both sides of transmon. Thus the couple constant at each leg can be either positive or negative. We derived the dynamics for both giant atom and bosonic field in the transmission in the harmonic limit. We calculated the scattering coefficients based on the linear response theory. We defined the non-Markovian regime for the giant atom from the property of scattering spectrum.
We explored the non-Markovianity of giant atom as a resource to manipulate the traveling field. Based on the bound state condition for the two-leg non-Markovian giant atom, we numerically simulated the process of catching and releasing the propagating bosonic field. We found the catching probability can be significantly enhanced using a superposition state of two counter propagating wave packets. In the end, we proposed a circuit-QED architecture to catch and release traveling microwave photon in the transmission line.

We outlook the problems and possible research directions for the future study. To increase further the catching probability, one possible solution is to use giant atom with  more than two coupling points with carefully design coupling strengths as the leaked field can be caught again by legs outside.
The shape of field pulse can also used to encode quantum information  but theory shows that a small atom with fixed coupling to a waveguide is only able to catch properly shaped traveling field, e.g., exponentially rising or  time-reversal symmetrized single-photon pulses \cite{li2022automatica}. Non-Markovian giant atom provides a possible solution to this problem. We would expect to generate and catch bound state with desired shape by designing the coupling constant at each leg  (possibly complex \cite{Wang2022qst}).
The nonlinear bound state condition  (\ref{resonantdark1}) in the non-Markovian regime also shows that one giant atom can catch two bound state simultaneously, which can be utilized to realize multimode quantum memories \cite{dai2023pra}.
For precise control of encoded quantum information in the flying qubit, it is also important to further examine the decoherence effects, speed, fidelity, and robustness of bound states.
Another interesting direction is to go beyond the single-excitation approximation and control multiphoton bound states for quantum information processing with bosonic code states \cite{Pfaff2017np}.

\section*{Acknowledgement}
This work was supported by
the NNSF of China (Nos.\ 11904261). The authors thank Florian Marquardt, G\"oran Johansson and  Anton Frisk Kockum for useful discussions.

\appendix

\section{\\Deviration for effective model} \label{AppA}
As mentioned in the main text, the giant atom with multiple legs divided into two separate groups can be simplified into an effective model, in which each group can be treat as a single coupling point with the effective coupling strength that can be tuned by the system parameters. Here, we present detailed derivations for the effective model from the EOM of atom.

We first discuss the case shown in Fig.~\ref{fig 2IDT}(a), where the  single-leg coupling strength is to be $c_{m_1}=(-1)^{m_1}c_1$  with  $m_1=1,2,\cdots,N_1$ and
$c_{m_2}=(-1)^{m_2}c_2$ with $m_2=N_1+1,\cdots, N_1+N_2$.
 If the
 bosonic field is initially in the ground state $\langle\hat{a}_k(0)\rangle=0$,  we have the EOM of
$\beta(t)\equiv\langle\hat{b}(t)\rangle$ from Eq.~(\ref{EOM2IDT}),
\begin{eqnarray}\label{APPEOM2IDT}
\frac{d}{dt}\beta(t)&=&-i\Omega\beta(t)-\Gamma
\sum_{m,m'}c_m c_{m'} \beta(t-|\tau_{mm'}|)\Theta(t-|\tau_{mm'}|)\nonumber\\
&=&(-i\Omega +N_1\gamma_1/2
+N_2\gamma
_2/2)\beta(t)\nonumber\\
&&-\sum_{j=1,2}
\gamma_j\sum_{n_j=0}^{N_j-1}(-1)^{n_j}(N_j-n_j)\beta(t-n_j\tau)\Theta(t-n_j\tau)\nonumber\\
&&-\sqrt{\gamma_1\gamma_2}\left[\sum_{n=0}^{N_1-1}(n+1)+\sum_{n=N_1}^{N_2-1}N_1
+\sum_{n=N_2}^{N_1+N_2-2}(N_1+N_2-n-1)\right]\nonumber\\
&&\times(-1)^{n+1}\beta(t-T-n\tau)\Theta(t-T-n\tau),
\end{eqnarray}
with parameters $\gamma_1=2\Gamma c_1^2$, and $\gamma_2=2\Gamma c_2^2$.
Note that the above EOM is written in the lab frame. In the RWA, the dynamics of $\beta(t)$ can be separated into a fast oscillation with frequency $\Omega$ and a low time evolution of amplitude $\beta(t)=\tilde{\beta}(t)e^{-i\Omega t }$.
Thus, for $t>0$, the EOM of $\tilde{\beta}(t)$ has the form
\begin{eqnarray}\label{EOMtilbeta}
\frac{d}{dt}\tilde{\beta}(t)
&=&( N_1\gamma_1/2
+N_2\gamma
_2/2)\tilde{\beta}(t)\nonumber\\
&&-\sum_{j=1,2}
\gamma_j\sum_{n_j=0}^{N_j-1}(-1)^{n_j}(N_j-n_j)e^{in_j\Omega  \tau}\tilde{\beta}(t-n_j\tau)\Theta(t-n_j\tau)\nonumber\\
&&-\sqrt{\gamma_1\gamma_2}\left[\sum_{n=0}^{N_1-1}(n+1)+\sum_{n=N_1}^{N_2-1}N_1
+\sum_{n=N_2}^{N_1+N_2-2}(N_1+N_2-n-1)\right]\nonumber\\
&&\times(-1)^{n+1}e^{i\Omega (T+n \tau)}\tilde{\beta}(t-T-n\tau)\Theta(t-T-n\tau).
\end{eqnarray}
If the time delays inside the two separated parts $N_1\tau$, $N_2\tau$ both are much shorter than the characteristic time of $\tilde\beta(t)$, we are allowed to neglect the delay time differences $n\tau$ in $\tilde{\beta}(t-n\tau)$, and replace the delay time differences $T+n\tau$ in $\tilde{\beta}(t-T-n\tau)$ by the traveling time between the centers of two separated parts $\tilde{T}=T+(N_1+N_2)\tau/2-\tau$, cf. the lower subfigure of Fig.~\ref{fig 2IDT}(a). Thus, we have
\begin{eqnarray}\label{EOMtilbeta}
\frac{d}{dt}\tilde{\beta}(t)
&\approx&\frac{N_1\gamma_1
+N_2\gamma
_2}{2}\tilde{\beta}(t)-\sum_{j=1,2}
\gamma_j\sum_{n_j=0}^{N_j-1}(-1)^{n_j}(N_j-n_j)e^{in_j\Omega\tau}\tilde{\beta}(t)\nonumber\\
&&-\sqrt{\gamma_1\gamma_2}\left[\sum_{n=0}^{N_1-1}(n+1)+\sum_{n=N_1}^{N_2-1}N_1
+\sum_{n=N_2}^{N_1+N_2-2}(N_1+N_2-n-1)\right]\nonumber\\
&&\times(-1)^{n+1}e^{i\Omega (T+n \tau)}\tilde{\beta}(t-\tilde{T})\Theta(t-\tilde{T})\nonumber\\
&=&\sum_{j=1,2}\gamma_j\left[\frac{N_j}{2}-
\sum_{n_j=0}^{N_j-1}(N_j-n_j)e^{in_j\phi}\right]\tilde{\beta}(t)\nonumber\\
&&+\sqrt{\gamma_1\gamma_2}\left[\sum_{n=0}^{N_1-1}(n+1)+\sum_{n=N_1}^{N_2-1}N_1 +\sum_{n=0}^{N_1-1}(N_1-n-1)e^{iN_2\phi}\right]\nonumber\\
&&\times e^{i\Omega T}e^{in\phi }\tilde{\beta}(t-\tilde{T})\Theta(t-\tilde{T})\nonumber\\
&=&\sum_{j=1,2}\gamma_j\left[\frac{N_j}{2}-
\sum_{n_j=0}^{N_j-1}(N_j-n_j)e^{in_j\phi}\right]\tilde{\beta}(t)\nonumber\\
&&+\sqrt{\gamma_1\gamma_2}e^{i\Omega T}(1-e^{iN_2\phi})\sum_{n=0}^{N_1-1}(n+1)e^{in\phi }\tilde{\beta}(t-\tilde{T})\Theta(t-\tilde{T})\nonumber\\
&&+N_1\sqrt{\gamma_1\gamma_2}e^{i\Omega T}\left[\sum_{n=N_1}^{N_2-1}e^{in\phi }+e^{iN_2\phi} \sum_{n=0}^{N_1-1}e^{in\phi } \right]\tilde{\beta}(t-\tilde{T})\Theta(t-\tilde{T})\nonumber\\
&=&-\sum_{j=1,2}\gamma_j\left[\frac{1-\cos
(N_j\phi)}{2(1-\cos \phi)}+i\frac{N_j\sin\phi-\sin (N_j\phi)}{2(1-\cos
\phi)}\right]\tilde{\beta}(t)\nonumber\\
&&+\sqrt{\gamma_1\gamma_2}\frac{(1-e^{iN_1\phi})(1-e^{iN_2\phi})}{(1-e^{i\phi})^2}e^{i\Omega T}\tilde{\beta}(t-\tilde{T})\Theta(t-\tilde{T})\nonumber\\
&=&-\sum_{j=1,2}\gamma_j\left[\frac{1-\cos
(N_j\phi)}{2(1-\cos \phi)}+i\frac{N_j\sin\phi-\sin (N_j\phi)}{2(1-\cos
\phi)}\right]\tilde{\beta}(t)\nonumber\\
&&+\sqrt{\gamma_1\gamma_2}\frac{\cos \Delta N \phi-\cos \bar{N}\phi }{1-\cos \phi }e^{i(\bar{N}-1)\phi}e^{i\Omega T}\tilde{\beta}(t-\tilde{T})\Theta(t-\tilde{T})\nonumber\\
&=&-i\tilde{\Delta}\tilde{\beta}(t)-\tilde{\gamma}_1\beta(t)-\tilde{\gamma}_2e^{i\Omega\tilde{T}}\tilde{\beta}(t-\tilde{T})\Theta(t-\tilde{T}),
\end{eqnarray}
where $\phi=\Omega\tau+\pi$. Here we have used the identity Eq.~(\ref{identity1}).
 With the even $N_1$ and $N_2$, we derive that the Lamb shift $\tilde{\Delta}$ and the effective decay rate $\tilde{\gamma}_1, \tilde{\gamma}_2$ as
\begin{eqnarray}
\tilde{\Delta}&=&-\frac{1}{2} \sum_{j=1,2}\gamma_j\frac{N_j\sin (\Omega \tau)+\sin(N_j\Omega\tau)}{1+\cos\Omega\tau},\label{EOM2IDT-2a}\\
\tilde{\gamma}_1&=&\frac{1}{2} \sum_{j=1,2}\gamma_j\frac{1-\cos(N_j\Omega\tau)}{1+\cos\Omega\tau},\label{EOM2IDT-2b}\\
\tilde{\gamma}_2&=&\sqrt{\gamma_1\gamma_2}\frac{\cos(\Delta N\Omega\tau)-\cos(\bar{N}\Omega\tau)}{1+\cos\Omega\tau},\label{EOM2IDT-2c}
\end{eqnarray}
where $\bar{N}=(N_1+N_2)/2$ and $\Delta N=(N_2-N_1)/2$.
Using the transformation $\tilde{\beta}(t)=e^{i\Omega t }\beta(t)$, we go back to the EOM of $\beta(t)$ given by
\begin{eqnarray}\label{AppEOM2IDT-1}
\frac{d}{dt}\beta(t)&=&-i(\Omega+\tilde{\Delta})\beta(t)
-\tilde{\gamma}_1\beta(t)-\tilde{\gamma}_2\beta(t-\tilde{T})\Theta(t-\tilde{T}).
\end{eqnarray}
Thus, we have proved that the
giant-atom structure with two separate groups of coupling points can be further simplified into an effective model of two coupling points.

We then discuss the case shown in Fig.~\ref{fig 2IDT}(b), where the single-leg coupling strength is set to be uniform, i.e., $c_{m_1}=c_1$  and
$c_{m_2}=c_2$ with  $m_1=1,2,\cdots,N_1$; $m_2=N_1+1,\cdots, N_1+N_2$.
The corresponding EOM of $\beta(t)$ is given by
\begin{eqnarray}\label{appEOM2IDTc}
\frac{d}{dt}\beta(t)
&=&(-i\Omega +N_1\gamma_1/2
+N_2\gamma
_2/2)\beta(t)\nonumber\\
&&-\sum_{j=1,2}
\gamma_j\sum_{n_j=0}^{N_j-1}(N_j-n_j)\beta(t-n_j\tau)\Theta(t-n_j\tau)\nonumber\\
&&-\sqrt{\gamma_1\gamma_2}\left[\sum_{n=0}^{N_1-1}(n+1)+\sum_{n=N_1}^{N_2-1}N_1
+\sum_{n=N_2}^{N_1+N_2-2}(N_1+N_2-n-1)\right]\nonumber\\
&&\times\beta(t-T-n\tau)\Theta(t-T-n\tau).
\end{eqnarray}
Based on the  derivation similar to Eq.~(\ref{EOMtilbeta}), the EOM of $\tilde{\beta}(t)$ has the form
\begin{eqnarray}\label{EOMtilbetac}
\frac{d}{dt}\tilde{\beta}(t)
&=&-\sum_{j=1,2}\gamma_j\left[\frac{1-\cos
(N_j\phi)}{2(1-\cos \phi)}+i\frac{N_j\sin\phi-\sin (N_j\phi)}{2(1-\cos
\phi)}\right]\tilde{\beta}(t)\nonumber\\
&&-\sqrt{\gamma_1\gamma_2}\frac{\cos \Delta N \phi-\cos \bar{N}\phi }{1-\cos \phi }e^{i(\bar{N}-1)\phi}e^{i\Omega T}\tilde{\beta}(t-\tilde{T})\Theta(t-\tilde{T}),
\end{eqnarray}
with parameter $\phi=\Omega\tau$. 
The EOM of $\beta(t)$
can also be expressed by the simplified model of Eq.~(\ref{AppEOM2IDT-1}), with the Lamb shift $\tilde{\Delta}$ and the effective decay rates $\tilde{\gamma}_1, \tilde{\gamma}_2$  given by
 \begin{eqnarray}
\tilde{\Delta}&=&\frac{1}{2}\sum_{j=1,2}\gamma_j\frac{N_j\sin (\Omega \tau)-\sin(N_j\Omega\tau)}{1-\cos\Omega\tau},\\
\tilde{\gamma}_1&=&\frac{1}{2} \sum_{j=1,2}\gamma_j\frac{1-\cos(N_j\Omega\tau)}{1-\cos\Omega\tau},\\
\tilde{\gamma}_2&=&\sqrt{\gamma_1\gamma_2}\frac{\cos(\Delta N\Omega\tau)-\cos(\bar{N}\Omega\tau)}{1-\cos\Omega\tau}.
\end{eqnarray}

\section*{References}


\begin{thebibliography}{10}
\expandafter\ifx\csname url\endcsname\relax
  \def\url#1{{\tt #1}}\fi
\expandafter\ifx\csname urlprefix\endcsname\relax\def\urlprefix{URL }\fi
\providecommand{\eprint}[2][]{\url{#2}}

\bibitem{Walls2008}
Walls D~F and Milburn G~J 2008 {\em {Quantum Optics}\/} 2nd ed (Springer) ISBN
  978-3-540-28573-1
  \urlprefix\url{http://link.springer.com/10.1007/978-3-540-28574-8}

\bibitem{anton2021book}
Frisk~Kockum A 2021 Quantum optics with giant atoms---the first five years {\em
  International Symposium on Mathematics, Quantum Theory, and Cryptography\/}
  ed Takagi T, Wakayama M, Tanaka K, Kunihiro N, Kimoto K and Ikematsu Y
  (Singapore: Springer Singapore) pp 125--146 ISBN 978-981-15-5191-8

\bibitem{Koch2007pra}
Koch J, Yu T~M, Gambetta J, Houck A~A, Schuster D~I, Majer J, Blais A, Devoret
  M~H, Girvin S~M and Schoelkopf R~J 2007 {\em Physical Review A\/} {\bf 76}
  042319 ISSN 1050-2947 (\textit{Preprint} \eprint{0703002})
  \urlprefix\url{http://link.aps.org/doi/10.1103/PhysRevA.76.042319}

\bibitem{Gustafsson2014science}
Gustafsson M~V, Aref T, Kockum A~F, Ekstr{\"{o}}m M~K, Johansson G and Delsing
  P 2014 {\em Science\/} {\bf 346} 207 ISSN 0036-8075 (\textit{Preprint}
  \eprint{1404.0401})
  \urlprefix\url{http://www.sciencemag.org/cgi/doi/10.1126/science.1257219}

\bibitem{Andersson2019np}
Andersson G, Suri B, Guo L, Aref T and Delsing P 2019 {\em Nature Physics\/}
  {\bf 15} 1123--1127 ISSN 1745-2481
  \urlprefix\url{https://doi.org/10.1038/s41567-019-0605-6}

\bibitem{Kannan2019}
Kannan B, Ruckriegel M, Campbell D, Kockum A~F, Braum{\"{u}}ller J, Kim D,
  Kjaergaard M, Krantz P, Melville A, Niedzielski B~M, Veps{\"{a}}l{\"{a}}inen
  A, Winik R, Yoder J, Nori F, Orlando T~P, Gustavsson S and Oliver W~D 2020
  {\em Nature\/} {\bf 583} 775 (\textit{Preprint} \eprint{1912.12233})
  \urlprefix\url{https://www.nature.com/articles/s41586-020-2529-9#citeas}

\bibitem{vadiraj2021pra}
Vadiraj A~M, Ask A, McConkey T~G, Nsanzineza I, Chang C~W~S, Kockum A~F and
  Wilson C~M 2021 {\em Phys. Rev. A\/} {\bf 103}(2) 023710
  \urlprefix\url{https://link.aps.org/doi/10.1103/PhysRevA.103.023710}

\bibitem{Wang2022nc}
Wang Z~Q, Wang Y~P, Yao J, Shen R~C, Wu W~J, Qian J, Li J, Zhu S~Y and You J~Q
  2022 {\em Nature Communications\/} {\bf 13} 7580 ISSN 2041-1723
  \urlprefix\url{https://doi.org/10.1038/s41467-022-35174-9}

\bibitem{Kockum2014pra}
Kockum A~F, Delsing P and Johansson G 2014 {\em Physical Review A\/} {\bf 90}
  013837 ISSN 1050-2947 (\textit{Preprint} \eprint{1406.0350})
  \urlprefix\url{http://link.aps.org/doi/10.1103/PhysRevA.90.013837}

\bibitem{Guo2017}
Guo L, Grimsmo A~L, Kockum A~F, Pletyukhov M and Johansson G 2017 {\em Physical
  Review A\/} {\bf 95} 053821 ISSN 2469-9926 (\textit{Preprint}
  \eprint{1612.00865})
  \urlprefix\url{http://link.aps.org/doi/10.1103/PhysRevA.95.053821}

\bibitem{Kockum2018prl}
Kockum A~F, Johansson G and Nori F 2018 {\em Physical Review Letters\/} {\bf
  120} 140404 ISSN 0031-9007 (\textit{Preprint} \eprint{1711.08863})
  \urlprefix\url{https://link.aps.org/doi/10.1103/PhysRevLett.120.140404}

\bibitem{carollo2020prr}
Carollo A, Cilluffo D and Ciccarello F 2020 {\em Phys. Rev. Res.\/} {\bf 2}(4)
  043184
  \urlprefix\url{https://link.aps.org/doi/10.1103/PhysRevResearch.2.043184}

\bibitem{guo2020prr}
Guo L, Kockum A~F, Marquardt F and Johansson G 2020 {\em Phys. Rev. Res.\/}
  {\bf 2}(4) 043014
  \urlprefix\url{https://link.aps.org/doi/10.1103/PhysRevResearch.2.043014}

\bibitem{taylor2020pra}
Guo S, Wang Y, Purdy T and Taylor J 2020 {\em Phys. Rev. A\/} {\bf 102}(3)
  033706 \urlprefix\url{https://link.aps.org/doi/10.1103/PhysRevA.102.033706}

\bibitem{Kockum2019b}
Kockum A~F 2019  (\textit{Preprint} \eprint{1912.13012})
  \urlprefix\url{http://arxiv.org/abs/1912.13012}

\bibitem{Gonzalez-Tudela2019}
Gonz{\'{a}}lez-Tudela A, {S{\'{a}}nchez Mu{\~{n}}oz} C and Cirac J~I 2019 {\em
  Physical Review Letters\/} {\bf 122} 203603 ISSN 0031-9007 (\textit{Preprint}
  \eprint{1901.00289})
  \urlprefix\url{https://link.aps.org/doi/10.1103/PhysRevLett.122.203603}

\bibitem{Calajo2019}
Calaj{\'{o}} G, Fang Y~L~L, Baranger H~U and Ciccarello F 2019 {\em Physical
  Review Letters\/} {\bf 122} 073601 ISSN 0031-9007 (\textit{Preprint}
  \eprint{1811.02582})
  \urlprefix\url{https://link.aps.org/doi/10.1103/PhysRevLett.122.073601}

\bibitem{Lorenzo2021sr}
Lorenzo S, Longhi S, Cabot A, Zambrini R and Giorgi G~L 2021 {\em Scientific
  Reports\/} {\bf 11} 12834 ISSN 2045-2322
  \urlprefix\url{https://doi.org/10.1038/s41598-021-92288-8}

\bibitem{sheremet2023rmp}
Sheremet A~S, Petrov M~I, Iorsh I~V, Poshakinskiy A~V and Poddubny A~N 2023
  {\em Rev. Mod. Phys.\/} {\bf 95}(1) 015002
  \urlprefix\url{https://link.aps.org/doi/10.1103/RevModPhys.95.015002}

\bibitem{zhao2020pra}
Zhao W and Wang Z 2020 {\em Phys. Rev. A\/} {\bf 101}(5) 053855
  \urlprefix\url{https://link.aps.org/doi/10.1103/PhysRevA.101.053855}

\bibitem{lim2023pra}
Lim K~H, Mok W~K and Kwek L~C 2023 {\em Phys. Rev. A\/} {\bf 107}(2) 023716
  \urlprefix\url{https://link.aps.org/doi/10.1103/PhysRevA.107.023716}

\bibitem{zhang2023pra}
Zhang X, Liu C, Gong Z and Wang Z 2023 {\em Phys. Rev. A\/} {\bf 108}(1) 013704
  \urlprefix\url{https://link.aps.org/doi/10.1103/PhysRevA.108.013704}

\bibitem{jia2023atomphoton}
Jia W~Z and Yu M~T 2023 Atom-photon dressed states in a waveguide-qed system
  with multiple giant atoms coupled to a resonator-array waveguide
  (\textit{Preprint} \eprint{2304.02072})

\bibitem{bag2023quantum}
Bag R and Roy D 2023 Quantum light-matter interactions in structured waveguides
  (\textit{Preprint} \eprint{2304.13306})

\bibitem{wang2023realizing}
Wang X, Zhu H~B, Liu T and Nori F 2023 Realizing quantum optics in structured
  environments with giant atoms (\textit{Preprint} \eprint{2304.10710})

\bibitem{chen2023giantatom}
Chen Y~T, Du L, Zhang Y, Guo L, Wu J~H, Artoni M and Rocca G~C~L 2023
  Giant-atom effects on population and entanglement dynamics of rydberg atoms
  (\textit{Preprint} \eprint{2304.14713})

\bibitem{cheng2022pra}
Cheng W, Wang Z and Liu Y~x 2022 {\em Phys. Rev. A\/} {\bf 106}(3) 033522
  \urlprefix\url{https://link.aps.org/doi/10.1103/PhysRevA.106.033522}

\bibitem{vega2021pra}
Vega C, Bello M, Porras D and Gonz\'alez-Tudela A 2021 {\em Phys. Rev. A\/}
  {\bf 104}(5) 053522
  \urlprefix\url{https://link.aps.org/doi/10.1103/PhysRevA.104.053522}

\bibitem{vega2023prr}
Vega C, Porras D and Gonz\'alez-Tudela A 2023 {\em Phys. Rev. Res.\/} {\bf
  5}(2) 023031
  \urlprefix\url{https://link.aps.org/doi/10.1103/PhysRevResearch.5.023031}

\bibitem{du2023giant}
Du L, Guo L, Zhang Y and Kockum A~F 2023 Giant emitters in a structured bath
  with non-hermitian skin effect (\textit{Preprint} \eprint{2308.16148})

\bibitem{Guimond2020}
Guimond P~O, Vermersch B, Juan M~L, Sharafiev A, Kirchmair G and Zoller P 2020
  {\em npj Quantum Information\/} {\bf 6} 32 ISSN 2056-6387
  \urlprefix\url{https://doi.org/10.1038/s41534-020-0261-9}

\bibitem{joshi2023prx}
Joshi C, Yang F and Mirhosseini M 2023 {\em Phys. Rev. X\/} {\bf 13}(2) 021039
  \urlprefix\url{https://link.aps.org/doi/10.1103/PhysRevX.13.021039}

\bibitem{zhang2021prx}
Zhang Y~X, i~Carceller C~R, Kjaergaard M and S\o{}rensen A~S 2021 {\em Phys.
  Rev. Lett.\/} {\bf 127}(23) 233601
  \urlprefix\url{https://link.aps.org/doi/10.1103/PhysRevLett.127.233601}

\bibitem{Kannan2023nat}
Kannan B, Almanakly A, Sung Y, Di~Paolo A, Rower D~A, Braum\"uller J, Melville
  A, Niedzielski B~M, Karamlou A, Serniak K, Veps\"al\"ainen A, Schwartz M~E,
  Yoder J~L, Winik R, Wang J~I~J, Orlando T~P, Gustavsson S, Grover J~A and
  Oliver W~D 2023 {\em Nature Physics\/} {\bf 19} 394--400 ISSN 1745-2481
  \urlprefix\url{https://doi.org/10.1038/s41567-022-01869-5}

\bibitem{Wang2022qst}
Wang X and Li H~R 2022 {\em Quantum Science and Technology\/} {\bf 7} 035007
  \urlprefix\url{https://dx.doi.org/10.1088/2058-9565/ac6a04}

\bibitem{du2022prl}
Du L, Zhang Y, Wu J~H, Kockum A~F and Li Y 2022 {\em Phys. Rev. Lett.\/} {\bf
  128}(22) 223602
  \urlprefix\url{https://link.aps.org/doi/10.1103/PhysRevLett.128.223602}

\bibitem{li2023tunable}
Li J, Lu J, Gong Z~R and Zhou L 2023 Tunable chiral bound states in a dimer
  chain of coupled resonators (\textit{Preprint} \eprint{2211.01734})

\bibitem{wang2021prl}
Wang X, Liu T, Kockum A~F, Li H~R and Nori F 2021 {\em Phys. Rev. Lett.\/} {\bf
  126}(4) 043602
  \urlprefix\url{https://link.aps.org/doi/10.1103/PhysRevLett.126.043602}

\bibitem{yin2023pra}
Yin X~L and Liao J~Q 2023 {\em Phys. Rev. A\/} {\bf 108}(2) 023728
  \urlprefix\url{https://link.aps.org/doi/10.1103/PhysRevA.108.023728}

\bibitem{yin2022pra}
Yin X~L, Luo W~B and Liao J~Q 2022 {\em Phys. Rev. A\/} {\bf 106}(6) 063703
  \urlprefix\url{https://link.aps.org/doi/10.1103/PhysRevA.106.063703}

\bibitem{santos2023prl}
Santos A~C and Bachelard R 2023 {\em Phys. Rev. Lett.\/} {\bf 130}(5) 053601
  \urlprefix\url{https://link.aps.org/doi/10.1103/PhysRevLett.130.053601}

\bibitem{zhu2022pra}
Zhu Y~T, Xue S, Wu R~B, Li W~L, Peng Z~H and Jiang M 2022 {\em Phys. Rev. A\/}
  {\bf 106}(4) 043710
  \urlprefix\url{https://link.aps.org/doi/10.1103/PhysRevA.106.043710}

\bibitem{Chen2022cp}
Chen Y~T, Du L, Guo L, Wang Z, Zhang Y, Li Y and Wu J~H 2022 {\em
  Communications Physics\/} {\bf 5} 215 ISSN 2399-3650
  \urlprefix\url{https://doi.org/10.1038/s42005-022-00991-3}

\bibitem{gu2023correlated}
Gu W, Huang H, Yi Z, Chen L, Sun L and Tan H 2023 Correlated two-photon
  scattering in a 1d waveguide coupled to two- or three-level giant atoms
  (\textit{Preprint} \eprint{2306.13836})

\bibitem{Ask2019}
Ask A, Ekstr{\"{o}}m M, Delsing P and Johansson G 2019 {\em Physical Review
  A\/} {\bf 99} 013840 ISSN 2469-9926 (\textit{Preprint} \eprint{1811.09421})
  \urlprefix\url{http://dx.doi.org/10.1103/PhysRevA.99.013840}

\bibitem{noachtar2022pra}
Noachtar D~D, Kn\"orzer J and Jonsson R~H 2022 {\em Phys. Rev. A\/} {\bf
  106}(1) 013702
  \urlprefix\url{https://link.aps.org/doi/10.1103/PhysRevA.106.013702}

\bibitem{zueco2022pra}
Terradas-Brians\'o S, Gonz\'alez-Guti\'errez C~A, Nori F, Mart\'{\i}n-Moreno L
  and Zueco D 2022 {\em Phys. Rev. A\/} {\bf 106}(6) 063717
  \urlprefix\url{https://link.aps.org/doi/10.1103/PhysRevA.106.063717}

\bibitem{Guimond2017}
Guimond P~O, Pletyukhov M, Pichler H and Zoller P 2017 {\em Quantum Science and
  Technology\/} {\bf 2} 044012 ISSN 2058-9565 (\textit{Preprint}
  \eprint{1706.07844})
  \urlprefix\url{http://dx.doi.org/10.1088/2058-9565/aa7f03}

\bibitem{cheng2023lp}
Cheng W, Wang Z and Tian T 2023 {\em Laser Physics\/} {\bf 33} 085203
  \urlprefix\url{https://dx.doi.org/10.1088/1555-6611/acde6e}

\bibitem{arranz2021prr}
Arranz~Regidor S, Crowder G, Carmichael H and Hughes S 2021 {\em Phys. Rev.
  Res.\/} {\bf 3}(2) 023030
  \urlprefix\url{https://link.aps.org/doi/10.1103/PhysRevResearch.3.023030}

\bibitem{pichler2016prl}
Pichler H and Zoller P 2016 {\em Phys. Rev. Lett.\/} {\bf 116}(9) 093601
  \urlprefix\url{https://link.aps.org/doi/10.1103/PhysRevLett.116.093601}

\bibitem{Wang2021oe}
Wang C, Ma X~S and Cheng M~T 2021 {\em Opt. Express\/} {\bf 29} 40116--40124
  \urlprefix\url{https://opg.optica.org/oe/abstract.cfm?URI=oe-29-24-40116}

\bibitem{Qiu2023}
Qiu Q~Y, Wu Y and L\"u X~Y 2023 {\em Science China Physics, Mechanics {\&}
  Astronomy\/} {\bf 66} 224212 ISSN 1869-1927
  \urlprefix\url{https://doi.org/10.1007/s11433-022-1990-x}

\bibitem{du2023pra}
Du L, Guo L and Li Y 2023 {\em Phys. Rev. A\/} {\bf 107}(2) 023705
  \urlprefix\url{https://link.aps.org/doi/10.1103/PhysRevA.107.023705}

\bibitem{ask2022prl}
Ask A and Johansson G 2022 {\em Phys. Rev. Lett.\/} {\bf 128}(8) 083603
  \urlprefix\url{https://link.aps.org/doi/10.1103/PhysRevLett.128.083603}

\bibitem{hannes2017pans}
Pichler H, Choi S, Zoller P and Lukin M~D 2017 {\em Proceedings of the National
  Academy of Sciences\/} {\bf 114} 11362--11367 (\textit{Preprint}
  \eprint{https://www.pnas.org/doi/pdf/10.1073/pnas.1711003114})
  \urlprefix\url{https://www.pnas.org/doi/abs/10.1073/pnas.1711003114}

\bibitem{Sletten2019}
Sletten L~R, Moores B~A, Viennot J~J and Lehnert K~W 2019 {\em Physical Review
  X\/} {\bf 9} 021056 ISSN 2160-3308 (\textit{Preprint} \eprint{1902.06344})
  \urlprefix\url{http://dx.doi.org/10.1103/PhysRevX.9.021056}

\bibitem{hann2019prl}
Hann C~T, Zou C~L, Zhang Y, Chu Y, Schoelkopf R~J, Girvin S~M and Jiang L 2019
  {\em Phys. Rev. Lett.\/} {\bf 123}(25) 250501
  \urlprefix\url{https://link.aps.org/doi/10.1103/PhysRevLett.123.250501}

\bibitem{divincenzo1995sci}
DiVincenzo D~P 1995 {\em Science\/} {\bf 270} 255--261 (\textit{Preprint}
  \eprint{https://www.science.org/doi/pdf/10.1126/science.270.5234.255})
  \urlprefix\url{https://www.science.org/doi/abs/10.1126/science.270.5234.255}

\bibitem{Barends2013}
Barends R, Kelly J, Megrant A, Sank D, Jeffrey E, Chen Y, Yin Y, Chiaro B,
  Mutus J, Neill C, O'Malley P, Roushan P, Wenner J, White T~C, Cleland A~N and
  Martinis J~M 2013 {\em Physical Review Letters\/} {\bf 111} 080502 ISSN
  0031-9007 (\textit{Preprint} \eprint{1304.2322})
  \urlprefix\url{http://link.aps.org/doi/10.1103/PhysRevLett.111.080502}

\bibitem{Rayleigh1885PLMS}
Rayleigh L 1885 {\em Proceedings of the London Mathematical Society\/} {\bf
  s1-17} 4--11 (\textit{Preprint}
  \eprint{https://londmathsoc.onlinelibrary.wiley.com/doi/pdf/10.1112/plms/s1-17.1.4})
  \urlprefix\url{https://londmathsoc.onlinelibrary.wiley.com/doi/abs/10.1112/plms/s1-17.1.4}

\bibitem{Datta1986SAW}
Datta S 1986 Surface acoustic wave devices
  \urlprefix\url{https://api.semanticscholar.org/CorpusID:137107497}

\bibitem{white2004book}
White R~M and Voltmer F~W 2004 {\em Applied Physics Letters\/} {\bf 7} 314--316
  ISSN 0003-6951 (\textit{Preprint}
  \eprint{https://pubs.aip.org/aip/apl/article-pdf/7/12/314/7521252/314\_1\_online.pdf})
  \urlprefix\url{https://doi.org/10.1063/1.1754276}

\bibitem{Peropadre2013}
Peropadre B, Lindkvist J, Hoi I~C, Wilson C~M, Garcia-Ripoll J~J, Delsing P and
  Johansson G 2013 {\em New Journal of Physics\/} {\bf 15} 035009 ISSN
  1367-2630 (\textit{Preprint} \eprint{1210.2264})
  \urlprefix\url{https://iopscience.iop.org/article/10.1088/1367-2630/15/3/035009}

\bibitem{Scully1997}
Scully M~O and Zubairy M~S 1997 {\em {Quantum Optics}\/} (Cambridge University
  Press) ISBN 9780521435956
  \urlprefix\url{https://www.cambridge.org/core/product/identifier/9780511813993/type/book}

\bibitem{Cai2021pra}
Cai Q~Y and Jia W~Z 2021 {\em Phys. Rev. A\/} {\bf 104}(3) 033710
  \urlprefix\url{https://link.aps.org/doi/10.1103/PhysRevA.104.033710}

\bibitem{Nakajima1982book}
Bilz H 1982 {\em Berichte der Bunsengesellschaft f\"ur physikalische Chemie\/}
  {\bf 86} 573--573 (\textit{Preprint}
  \eprint{https://onlinelibrary.wiley.com/doi/pdf/10.1002/bbpc.19820860623})
  \urlprefix\url{https://onlinelibrary.wiley.com/doi/abs/10.1002/bbpc.19820860623}

\bibitem{Kubo1957JPSJ}
Kubo R 1957 {\em Journal of the Physical Society of Japan\/} {\bf 12} 570--586
  (\textit{Preprint} \eprint{https://doi.org/10.1143/JPSJ.12.570})
  \urlprefix\url{https://doi.org/10.1143/JPSJ.12.570}

\bibitem{du2021prr}
Du L, Chen Y~T and Li Y 2021 {\em Phys. Rev. Res.\/} {\bf 3}(4) 043226
  \urlprefix\url{https://link.aps.org/doi/10.1103/PhysRevResearch.3.043226}

\bibitem{li2022prb}
Li W, Dong X, Zhang G and Wu R~B 2022 {\em Phys. Rev. B\/} {\bf 106}(13) 134305
  \urlprefix\url{https://link.aps.org/doi/10.1103/PhysRevB.106.134305}

\bibitem{li2022automatica}
Li W~L, Zhang G and Wu R~B 2022 {\em Automatica\/} {\bf 143} 110338 ISSN
  0005-1098
  \urlprefix\url{https://www.sciencedirect.com/science/article/pii/S000510982200187X}

\bibitem{yin2013prl}
Yin Y, Chen Y, Sank D, O'Malley P~J~J, White T~C, Barends R, Kelly J, Lucero E,
  Mariantoni M, Megrant A, Neill C, Vainsencher A, Wenner J, Korotkov A~N,
  Cleland A~N and Martinis J~M 2013 {\em Phys. Rev. Lett.\/} {\bf 110}(10)
  107001
  \urlprefix\url{https://link.aps.org/doi/10.1103/PhysRevLett.110.107001}

\bibitem{pierre2014apl}
Pierre M, Svensson I~M, Raman~Sathyamoorthy S, Johansson G and Delsing P 2014
  {\em Applied Physics Letters\/} {\bf 104} 232604 ISSN 0003-6951
  (\textit{Preprint}
  \eprint{https://pubs.aip.org/aip/apl/article-pdf/doi/10.1063/1.4882646/13940566/232604\_1\_online.pdf})
  \urlprefix\url{https://doi.org/10.1063/1.4882646}

\bibitem{dai2023pra}
Dai G, He K, He Y, Zhao C, Liu J and Chen W 2023 {\em Phys. Rev. A\/} {\bf
  107}(4) 042428
  \urlprefix\url{https://link.aps.org/doi/10.1103/PhysRevA.107.042428}

\bibitem{Pfaff2017np}
Pfaff W, Axline C~J, Burkhart L~D, Vool U, Reinhold P, Frunzio L, Jiang L,
  Devoret M~H and Schoelkopf R~J 2017 {\em Nature Physics\/} {\bf 13} 882--887
  ISSN 1745-2481 \urlprefix\url{https://doi.org/10.1038/nphys4143}

\end{thebibliography}
\bibliographystyle{iopart-num}

\providecommand{\newblock}{}

\end{document}